\documentclass[12pt]{article}
\usepackage{setspace}
\usepackage[margin=1in]{geometry}

\usepackage{mathptmx}
\usepackage[T1]{fontenc}
\usepackage{comment}

\usepackage{siunitx}
\usepackage[bb=boondox]{mathalfa}
\usepackage{bbm,amsmath,amssymb}

\usepackage[labelfont=bf]{caption}
\usepackage[labelsep=period]{caption}

\usepackage{float}

\usepackage{hyperref}

\usepackage{longtable}

\usepackage{titlesec}
\titleformat*{\section}{\Large \bfseries}
\titleformat*{\subsection}{\large \itshape}
\titleformat*{\subsubsection}{\itshape}

\usepackage{natbib}
\usepackage{color}
\usepackage{graphicx,subcaption}

\newcommand{\hl}[1]{{#1}}

\title{\textbf {\Large Parameter Estimation Procedures for Exponential-family Random Graph Models on Count-valued Networks: A Comparative Simulation Study\footnote{Manuscript accepted by \textit{Social Networks}: \href{https://doi.org/10.1016/j.socnet.2023.07.001}{https://doi.org/10.1016/j.socnet.2023.07.001}}
}}

\author{Peng Huang\thanks{Departments of Sociology, and Statistics, University of California, Irvine}\and Carter T. Butts\footnotemark[2]~ \thanks{Departments of Computer Science, and EECS, University of California, Irvine. Email: \texttt{buttsc@uci.edu}}}
\date{}
\begin{document}

\maketitle

\noindent \textbf{\large Abstract}

\noindent The exponential-family random graph models (ERGMs) have emerged as an important framework for modeling social networks for a wide variety of relational types.  ERGMs for valued networks are less well-developed than their unvalued counterparts, and pose particular computational challenges. Network data with edge values on the non-negative integers (count-valued networks) is an important such case, with examples ranging from the magnitude of migration and trade flows between places to the frequency of interactions and encounters between individuals. Here, we propose an efficient parallelable subsampled maximum pseudo-likelihood estimation (MPLE) scheme for count-valued ERGMs, and compare its performance with existing Contrastive Divergence (CD) and Monte Carlo Maximum Likelihood Estimation (MCMLE) approaches via a simulation study based on migration flow networks in two U.S. states. Our results suggest that edge value variance is a key factor in method performance, while network size mainly influences their relative merits in computational time. For small-variance networks, all methods perform well in point estimations while CD greatly overestimates uncertainties, and MPLE underestimates them for dependence terms; all methods have fast estimation for small networks, but CD and subsampled multi-core MPLE provides speed advantages as network size increases. For large-variance networks, both MPLE and MCMLE offer high-quality estimates of coefficients and their uncertainty, but MPLE is significantly faster than MCMLE; MPLE is also a better seeding method for MCMLE than CD, as the latter makes MCMLE more prone to convergence failure.
The study suggests that MCMLE and MPLE should be the default approach to estimate ERGMs for small-variance and large-variance valued networks, respectively. We also offer further suggestions regarding choice of computational method for valued ERGMs based on data structure, available computational resources and analytical goals.

\medskip
\noindent \textbf{\large Keywords}

\noindent Contrastive divergence, Exponential-family random graph model, Markov chain Monte Carlo, Maximum likelihood estimation, Pseudo likelihood, Valued/Weighted networks

\section{Introduction}
Binary relations - relations in which edges can be approximated as simply ``present'' or ``absent'' - form the backbone of the social network field, with decades of theoretical, methodological, and empirical progress in understanding their structure and function.  Valued relations, while by no means neglected, are less well-understood, and our tools for studying them less well-developed. Yet, the ``strength of social ties'' is at core of many scientific questions in a range of social settings \citep{granovetter_strength_1973,mcmillan2022worth}. Examples of valued relations include the frequency of interaction in interpersonal contact networks \citep{bernard.et.al:sn:1979}, number of cosponsored bills shared among legislators \citep{cranmer_inferential_2011,fowler2006connecting}, communication volume within and among organizations \citep{drabek.et.al:bk:1981,butts.et.al:jms:2007}, encounters among non-human animals \citep{faust2011animal}, and trade and migration flows among nations \citep{windzio_network_2018,ward_gravitys_2013}.
 The need for rich information about social relations is particularly acute for networks involving interactions among aggregate entities such as nations, geographical areas, gangs, or formal organizations: because ties in such networks are themselves frequently aggregations of lower-level interactions, it is often the case that one's interest is not in the mere existence of trade, migration, homicide, communication, or other interactions, but their volume, frequency, or other quantitative features. In such settings, modeling edge states is of considerable substantive importance.

The earliest statistical modeling of valued relations was accomplished via network regression methods \citep{krackhardt:sn:1988}; these provide only least-squares estimates of covariate effects, although autocorrelation-robust null hypothesis tests for such effects are well-known \citep{dekker.et.al:p:2007}, and some generalization via generalized linear models (GLMs) and related techniques is possible.  Some forms of dependence can, further, be controlled semi-parametrically using latent structure models \citep[e.g., ][]{nowicki.snijders:jasa:2001,hoff.et.al:jasa:2002,vu.et.al:aas:2013,aicher.et.al:jcn:2014}, allowing estimation of covariate effects while accounting for unobserved mechanisms that can be written in terms of mixing on unobserved variables.  Parametric models for valued graphs with general classes of dependence effects have been longer in coming, the current state of the art being exponential family random graph models (ERGMs) defined on sets of valued graphs \citep{block2022statistical,desmarais_statistical_2012,krivitsky_exponential-family_2012, krivitsky_exponential-family_2017}; but see also \citet{robins.et.al:p:1999} for a pioneering example using categorical data and pseudo-likelihood estimation).  Although ERGMs for valued graphs are not complete in the sense that they are for unvalued graphs (i.e., for most types of edge values, it is not always possible to write an arbitrary distribution on the order-$N$ valued graphs in ERGM form), they are still highly general families, able to flexibly specify a wide range of effects.  Since their introduction, they have been applied in a number of settings, ranging from networks of collaboration in government, and networks of migration flows, to networks of functional connectivity between brain regions \citep{huang2022rooted,simpson_analyzing_2013,ulibarri_linking_2017,windzio_network_2018}.

Notwithstanding their broad applicability, parameter estimation for ERGMs in practice can be computationally demanding, a problem that is especially acute for valued networks. This issue has clearly had an impact on empirical network analyses in the published literature, forcing researchers to employ compromises or workarounds. As an example, \cite{aksoy_model_2020} noted in their paper that they could not obtain convergence for a single 81-node network using valued ERGMs. For research that managed to obtain ERGM estimation of their valued networks, they had to either dichotomize the data and fall back to binary models \citep{leal_network_2021}, or coarsen the counts into quintiles \citep{windzio_network_2018,windzio_network_2019}; data transformation of this type greatly reduces computational difficulties, but in the meantime brings information loss and underestimation of variability \citep{altman2006cost}. In short, even though methodological advances in valued network modeling have made it possible for researchers to capture quantitative features of relations beyond dichotomizational operations \citep{cranmer_inferential_2011}, the computational load remains a lingering hurdle to fully exploit the potential of these methods in scientific applications.

The major computational cost of ERGM estimation comes from the normalizing factor in its likelihood function, which is generally an intractable function involving the sum or integral of an exponentiated potential over the set of all possible network configurations.  Although much is made over the fact that these sums have too many elements to explicitly evaluate \citep[except in the case of extremely small unvalued graphs, e.g. ][]{von.et.al:sn:2021}, this is not the major obstacle to computation: rather, the difficulty rises from the extreme roughness (i.e., high variance) of the exponentiated potential over the support, which (in the absence of an explicit analytical solution) renders naive attempts at numerical approximation ineffective.  This problem can be amplified in the valued case, particularly where edge values vary greatly; valued edges can also pose challenges for some approximate estimation procedures that are successful in the case of unvalued ERGMs, as they must now explore a larger \emph{per edge} state space.  This high cost of estimation puts a priority on computationally efficient approximation methods.  However, there has not been to date a systematic study of how well such methods perform, either in terms of improved computational efficiency or quality of estimation.

This paper provides a look at this issue, evaluating estimation quality and computational cost for a number of alternative valued ERGM estimation techniques.  We focus on ERGMs for count-valued networks, i.e. relations whose edges take values on the unbounded non-negative integers, evaluating estimators via a simulation study based on intercounty migration-flow networks in two U.S. states. We vary the variance of edge values and the (node) size of the network, to simulate different data structures. The methods examined include the two currently implemented ``standard'' strategies - contrastive divergence \citep[CD; ][]{krivitsky_using_2017} and Markov Chain Monte Carlo maximum likelihood estimation \citep[MCMLE; ][]{hunter2012computational} - as well as one approach not previously used in this setting, maximum pseudo-likelihood estimation (MPLE). MPLE is a workhorse approximation method in the binary ERGM case \citep{strauss_pseudolikelihood_1990}, but requires special implementation measures for the count-data case, and to our knowledge has not previously been used for count-data ERGMs with general dependence. We also compare the performance of MPLE and CD as two seeding options for MCMLE. For all methods, we evaluate their computational speed, bias, variability, accuracy, calibration of estimated standard errors and confidence coverage.

The remainder of the paper proceeds as follows. Section~\ref{sec_ergm} briefly reviews ERGMs for valued networks, with applicable estimation strategies discussed in Section~\ref{sec_estmeth}.  Our simulation study design is described in Section~\ref{sec_design}, with results reported in Section~\ref{sec_results}.  Section~\ref{sec_disc} discusses implications for method selection, and Section~\ref{sec_conc} concludes the paper.

\section{Count-valued ERGMs} \label{sec_ergm}

An ERGM family for count data can be written as,
\begin{equation} \label{eq:main}
\Pr(Y=y|\theta,X)=\frac{h(y)\exp(\theta^T g(y,X))}{\sum_{y' \in \mathcal{Y}} h(y') \exp(\theta^T g(y',X))},
\end{equation}
\noindent where $y$ is a realization of the network random variable $Y$ on support $\mathcal{Y}$, the elements of which are graphs whose edges take values on the set $\{0,1,\ldots\}$.  (Here, we further assume that $\mathcal{Y}$ is a subset of the order-$N$ count-valued graphs, though generalization is possible.)  $g: \mathcal{Y},X \mapsto \mathbb{R}^k$ is a vector of sufficient statistics, determined by exogenous covariates $X$ and the graph state $y$, with corresponding parameter vector $\theta$.  Finally, $h: \mathcal{Y} \mapsto \mathbb{R}_{\ge 0}$ is the \emph{reference measure}, which defines the limiting behavior of the model as $\theta \to 0$.  Often tacitly taken to be constant for binary ERGMs, the reference measure is essential for valued ERGMs, as it determines the marginal distribution of edge values under the reference \citep{krivitsky_exponential-family_2012}.  Leaving $h(y) \propto 1$ leads to a marginal Boltzmann baseline distribution, while choosing
\begin{equation}
h(y)=\prod_{(i,j) \in \mathbb{Y}}{(y_{ij}!)}^{-1}
\end{equation}
\noindent where $\mathbb{Y}$ is the set of edge variables, and $y_{ij}$ is the value of the $(i,j)$ edge, leads to a marginal Poisson baseline distribution of edge values.  Other choices are also possible, some of which may have specific substantive interpretations \citep[see e.g.][for examples in the binary case]{butts:jms:2019,butts:jms:2020a}.

As with binary ERGMs, we may specify the conditional probability that a given $i,j$ edge variable will take a specified value.  Again interpreting $Y$ and $y$ as random adjacency matrices, let $Y^c_{ij}$ (respectively $y^c_{ij}$) refer to the set of all edge variables other than the $ij$th, and let the notation $z \cup Y^c_{ij}$ refer to the network formed by $Y$ with the $ij$th edge variable set to value $z$.  Then we have
\begin{align}
\Pr(Y_{ij}=y_{ij}|Y^c_{ij}=y^c_{ij},\theta,X) &= \frac{h(y_{ij}\cup y^c_{ij}) \exp(\theta^T g(y_{ij}\cup y^c_{ij},X))}{\sum_{\ell=0}^{\infty} h(\ell \cup y^c_{ij}) \exp(\theta^T g(\ell \cup y^c_{ij},X))}  \label{eq:condprob}\\
&= \left[ \sum_{\ell=0}^{\infty} \frac{h(\ell \cup y^c_{ij})}{h(y_{ij} \cup y^c_{ij})} \exp\left[\theta^T \left(g(\ell \cup y^c_{ij},X) - g(y_{ij} \cup y^c_{ij},X) \right)\right] \right]^{-1}. \label{eq:condprob2}
\end{align}

While the derivation is identical to the binary case (as can be appreciated by noting that Eq.~\ref{eq:condprob} would reduce to the usual logistic form if $\ell$ were restricted to be $\le 1$), we note the computationally important difference that the conditional edge probability itself now has a non-trivial normalizing factor.  In the general case, this has no analytical solution, and since it involves an infinite sum it cannot be explicitly evaluated otherwise.  Although this does not impact e.g. the acceptance calculations for typical Markov Chain Monte Carlo (MCMC) algorithms (since the conditional odds of one graph versus another does not depend upon either normalizing factor), it does affect computation for the MPLE (which does depend on the conditional edge probability).  Here, we formulate a finite sum approximation to Eq.\ref{eq:condprob} for MPLE, as described below.

\section{Estimation strategies for count-valued ERGMs } \label{sec_estmeth}
While many approaches to parameter estimation are possible, we focus here on approximations to the maximum likelihood estimator (MLE).  Here, we briefly review the strategies employed, including special considerations for the count-data case.  We note that some alternative schemes explored in the binary case (e.g., variational methods \citep{mele2017structural,tan2020bayesian,wainwright2008graphical}) may be adapted to the count data problem, but for purposes of this paper we limit our study to approaches that have been established as broadly useful, and (with the exception of MPLE, which we extend) for which count-valued implementations currently exist.

\subsection{Monte Carlo Maximum Likelihood Estimation}

There are currently two widely used schemes for MCMC-based maximum likelihood estimation: stochastic approximation \citep{snijders_markov_2002,wang_pnet_2009}, which is based on attempting to match the expected sufficient statistics to their observed values (exploiting the coincidence of methods-of-moments and MLE for exponential families); and the Geyer-Thompson method  \citep{geyer_constrained_1992,hunter_ergm_2008} (supplemented in current implementations by Hummel stepping \citep{hummel_improving_2012}), which uses an importance sampling scheme to directly optimize the log-likelihood surface.  We here employ the former in its \texttt{statnet} implementation \citep{hunter_ergm_2008,krivitsky_package_2012}.  

MCMLE methods are the current gold-standard techniques for ERGM maximum likelihood estimation, with good theoretical properties \citep{snijders_markov_2002,handcock:ch:2003} and strong performance in simulation studies for binary networks \citep{van_duijn_framework_2009}.  An important bottleneck impacting the use of MCMLE, however, is the ability to produce relatively high-quality draws from the specified ERGM distribution (without which, the algorithms will not converge correctly).  While it is known that conventional MCMC algorithms can in principle mix arbitrarily slowly \citep{snijders_markov_2002,bhamidi.et.al:aap:2011}, in actual practice this problem has been observed primarily in badly specified models that are degenerate or near-degenerate, and hence of limited relevance in typical social network applications \citep{hunter2012computational}. That said, estimation time can still become long on very large networks, particularly for models with strong edgewise dependence.  

This cost issue becomes more acute for valued ERGMs, especially where edge values are highly variable.  Intuitively, good MCMC mixing requires the Markov chain to explore the space of high-probability graphs, whose size increases substantially when edge values vary over a large range.  For instance, for a simple random walk MCMC algorithm that proposes perturbing edges at random,\footnote{Practical implementations often use slightly different proposals, but the basic intuition carries.} $\mathcal{O}(N^2)$ toggles may be needed to ensure that every edge variable in an unvalued graph has a high probability of having the ``opportunity'' to change state.  If edges typically vary over some interval of order $R$, then a similar random walk scheme that increments or decrements edge values will need at least $\mathcal{O}(R {N^2})$ for each edge to have the ``opportunity'' to cover its range of values.  For networks with large counts (e.g., migrant-flow networks), one can easily obtain $R\gg N$, in which case simulation costs can rapidly become prohibitive.  Although this problem can be alleviated by coarsening edge values to a much smaller range \citep[as done e.g. by][]{windzio_network_2018, windzio_network_2019}, this both loses information and distorts the resulting model (since e.g., coarsening artificially reduces the entropy of the graph distribution).  In principle, improved MCMC algorithms offer a better way to address this problem in the long-term, but  current implementations do not seem to scale well for high-variance count models \citep[e.g.][]{aksoy_model_2020}.  As we show below, MPLE can often deliver comparable estimation quality to MCMLE for high-$R$ valued ERGMs, where the latter suffers substantial increases in computational cost.

\subsection{Contrastive Divergence}

One alternative to either numerical approximation of expected statistics or of log likelihood ratios is to use a local approximation to the gradient of the likelihood in regions of the support ``near'' the observed data.  This is the essential idea behind contrastive divergence (CD) \citep{hinton_training_2002}, a method originally introduced in the machine learning literature for scalable inference that is particularly well-suited to ERGMs and other exponential families \citep{ascuncion.et.al:icml:2010,krivitsky_using_2017}.  CD can be employed for both valued and unvalued graphs, and greatly reduces computational time by using using only very short MCMC chains starting at the observed data, depending on neither sample convergence nor burn-in.  It is, however, an approximate technique that optimizes a function closely related to the pseudo-likelihood \citep{ascuncion.et.al:icml:2010}, and thus shares some of the drawbacks of the MPLE. These properties make CD a reasonable seeding method that offers MCMLE with starting values for estimators, as starting values close to the MLE is known to help reduce iteration rounds and avoid convergence failures for MCMC algorithms. \cite{krivitsky_using_2017} found that MPLE typically outperformed CD as a seeding method for MCMLE in the binary ERGM regime; but since MPLE has not yet been implemented for valued ERGMs, CD currently serves as the default seeding method for MCMLE in the \texttt{statenet} package for valued graphs. Here, we evaluate CD both as a standalone method and a seeding method, in comparison with MPLE, for MCMLE.

\subsection{Maximum Pseudo-Likelihood Estimation} \label{sec_mple}

Although maximum pseudo-likelihood
estimation (MPLE) has not to our knowledge been studied or implemented for count-valued ERGMs, it is an otherwise well-known technique (being the first practical method for general ERGM estimation \citep{strauss_pseudolikelihood_1990}). MPLE optimizes the product of the conditional likelihoods of each edge variable (the eponymous pseudo-likelihood \citep{besag_spatial_1974}).  In the unvalued case, this reduces to a logistic regression problem, allowing the MPLE to be obtained using standard regression algorithms \citep[a fact that was instrumental in its early adoption, see e.g.][]{c.anderson.et.al:sn:1999}. The MPLE is known to be consistent in some asymptotic scenarios \citep{hyvarinen2006consistency,strauss_pseudolikelihood_1990}.  For finite scenarios, in the special case of edgewise independent ERGMs, the MPLE coincides with the MLE; this ceases to be true for dependence models, though the MPLE is generally close enough to the MLE to be used as a standard method for initializing MCMLE estimators, and its first-order performance on large networks can be very good \citep{an_fitting_2016, schmid_exponential_2017}.  Because it does not fully account for interactions among edge variables, the pseudo-likelihood function tends to be excessively concentrated, leading to poor calibration of standard error estimates \citep[as shown in binary ERGMs:][]{lubbers_comparison_2007,van_duijn_framework_2009}.  However, MPLE computation can be quite efficient, further aided by the fact that (1) the pseudo-likelihood itself can be approximated by subsampling edge variables, rather than computing on all of them, and (2) the calculations in question are embarrassingly parallel, making it possible to greatly reduce wall-clock time on multi-core CPUs. %

As noted above, MPLE computation in the count-data context is more complex than in the binary case, and to our knowledge it has not been previously studied for count-valued ERGMs with dyadic dependence.  We thus consider it here in greater detail.  As in the binary case, the MPLE is defined by

\begin{equation}
\hat{\theta}_{\mathrm{MPLE}} = \arg\max_\theta \prod_{(i,j) \in \mathbb{Y}} \Pr(Y_{ij}=y_{ij}|Y^c_{ij}=y^c_{ij},\theta,X), \label{eq:mple}
\end{equation}

\noindent where the conditional probabilities in question are given by Eq.~\ref{eq:condprob} and~\ref{eq:condprob2}.  Per Eq.~\ref{eq:condprob2}, these latter conditionals depend upon a sum over the possible edge states of products of two factors: one involves the ratio of the reference measure at the observed edge value versus its alternative values, and the other involves the exponentiated difference in sufficient statistics between the observed network and the same network with the focal edge taking on alternative values.  For the former, we observe that (in the case of the Poissonian reference), we have
\begin{align}
\frac{h(\ell \cup y^c_{ij})}{h(y_{ij} \cup y^c_{ij})} &= \frac{ (\ell!)^{-1 }\prod_{(k,l) \in \mathbb{Y}\setminus (i,j)}{(y_{kl}!)}^{-1} }{ (y_{ij}!)^{-1 }\prod_{(k,l) \in \mathbb{Y}\setminus (i,j)}{(y_{kl}!)}^{-1} } \nonumber\\
&= \frac{y_{ij}!}{\ell!}, \label{eq:hrat}
\end{align}
while the latter is simply
\[
\exp\left[\theta^T \Delta_{ij}(y,\ell)\right],
\]
where $\Delta$ is the ``generalized'' changescore
\[
\Delta_{ij}(y,\ell) = g(\ell \cup y^c_{ij},X) - g(y_{ij} \cup y^c_{ij},X).
\]
There is not, in general, a simple form for the sum of these terms over all $\ell$.  However, we observe that the ratio of Eq.~\ref{eq:hrat} falls very rapidly (as roughly $\ell^{-\ell}$) for $\ell \gg y_{ij}$, and it is hence possible in practice to approximate the infinite sum by truncation.  More generally, we employ several techniques for improving computational performance, as described in the following subsections.

\subsubsection{Pre-caching of ratios and differences} We note that neither the ratio of reference measures nor the changescores depend upon $\theta$.  Considerable computational savings can hence be had by pre-computing the ratios of Eq.~\ref{eq:hrat} and the $\Delta$ values for the necessary range of $\ell$ values on each edge.  This carries a storage cost that scales with the product of the $\ell$ range and the number of edge variables used, but avoids frequent recalculation of these (expensive) quantities on each pseudo-likelihood evaluation.

\subsubsection{Edge sum truncation and/or coarsening}\label{sec_truncation} Per Eq.~\ref{eq:condprob}, the conditional log-likelihood of each edge variable involves as sum over $\ell\in 0,\ldots,\infty$.  As noted above, we may approximate this sum by instead evaluating it over $\ell\in 0,\ldots,\ell_{\max}$, where $\ell_{\max}$ is large enough to be dominated by the decline in $y_{ij}!/\ell!$.  Where the marginal distributions of each edge variable can be approximated as roughly Poissonian, choosing $\ell_{\max} = \lambda \max_{(i,j) \in \mathbb{Y}} y_{ij}$ with e.g. $\lambda\approx 4$ is an extremely conservative approach.  (This is based on the observation that a Poisson random variable with expectation $z\geq2$ has a 99.9\% quantile for $4z$, assuming conservatively the expectation is the maximum observed value). $\lambda$ can be further reduced as its max observed value increases (because the quantile of $\lambda z$ grows with $z$).

Truncation using the above method is adequate for small networks, or networks with low edge counts.  However, when edge counts become extremely large, considerable computational effort may be wasted in computing conditional probabilities for small $\ell$ values when the observed value is large, or for large $\ell$ when the observed value is small. Using the same Poissonian approximation, we may further improve performance by working with the edgewise doubly-truncated sum over $\ell \in \ell^{ij}_{\min},\ldots,\ell^{ij}_{\max}$, with $\ell^{ij}_{\min} = \max[0, y_{ij} - 4\lambda \sqrt{y_{ij}}]$ and $\ell^{ij}_{\max} = y_{ij} + 4\lambda \sqrt{y_{ij}}$.  Because it is common to have network effects that can strongly suppress edges, however, we also recommend retaining some very small edge values as a buffer. Valued social networks usually have right-skewed distributions of edge values, so adding a few small edge values can also effectively cover the empirical distribution without significant increase in computation load. Our code by default retains integers from 0 to 5, although we strongly encourage extending coverage to the closest integer of sample mean of $y$, $[\mu]$ when feasible. This also defines the support of edges whose observed value is zero. The approach then leads to sums over $\ell$ values of the form $\ell \in \{0,\ldots, [\mu] \} \cup \{\ell^{ij}_{\min},\ldots,\ell^{ij}_{\max}\}$.  This usually retains $\mathcal{O}(\sqrt{y_{ij}})$ terms per sampled edge, which is often a substantial savings as $y_{ij}$ values become large.

When dealing with extremely large counts, storing and computing even $\sqrt{y}$ terms can become prohibitive (particularly if many edge variables are needed for adequate statistical power).  In such cases, a coarsened approximation to the sum is another option.  To coarsen, we select $k$ evenly spaced values from $\ell^{ij}_{\min},\ldots,\ell^{ij}_{\max}$, and compute the associated contributions to the edge sum only for these terms.  We also include, however, the gap (in terms of the number of ``skipped'' $\ell$ values) between subsequent calculated terms, and weight each computed term by the number of elements in the gap; this is equivalent to approximating the sum via a step function, with knots at the computed $\ell$ values.  Our experience with this method has been promising, although problems can ensue if the sum becomes heavily concentrated on a range of terms that lie within adjacent knots. We thus do not employ this technique in this paper, although we offer it as a promising target for future research.  Of course, other approximation methods are also possible (e.g., integral approximations), and may be useful in the large-count regime.

\subsubsection{Edge variable sampling} Although the exact calculation of the pseudo-likelihood is at least $\mathcal{O}(R{N^2})$, the log pseudo-likelihood can easily be approximated by random sampling of edge variables; this reduces both storage and computational cost. As shown by our experiments, subsampled MPLE can yield high-quality estimates with less time consumed. Our implementation offers different sampling schemes such as uniform random sampling, as well as weighted (i.e., importance) sampling schemes analogous to the ``Tie-No Tie'' proposal method frequently used in ERGM MCMC \citep{morris_specification_2008}. The two schemes are almost identical in our study case because the binary density is close to 0.5, and we use the random sampling scheme in this paper for simplicity.

\subsubsection{Parallel evaluation of conditional log-likelihoods} Because the log of the pseudo-likelihood is linearly separable, its calculation is an embarrassingly parallel problem.  In practice, we divide sampled edge variables into batches, and calculate their conditional log-likelihoods independently on different cores.  %
This leads to wall-clock time reductions, as the pseudo-likelihood calculation time scales with the inverse of the number of available cores.  This (combined with edge variable sampling) can make the MPLE an attractive choice for very large valued networks, especially when many cores are available.

Taken together, the above computational techniques allow the MPLE to be used even for very large networks with highly dispersed counts (although not all of them are needed when counts are less variable, or on smaller networks).  As we show, valued MPLE is very fast, and the resulting estimator can have low bias and high accuracy for valued networks; it offers high-quality calibration of uncertainty when the edge variance is large, but is prone to overconfidence for dependence terms (i.e., underestimation of the standard error) and conservative for nondependence terms when the edge variance is small.

\section{Study design} \label{sec_design}

We evaluate the above estimation techniques via a parameter recovery study, in which we generate networks from a realistic generative model based on an initial fit to real-world social networks, estimate models to the simulated draws using each respective technique, and then examine the properties of the resulting estimators.  Our generative model was created by fitting a valued ERGM to an empirical case (see below) using MCMLE; we then obtained 500 high-quality draws from the fitted ERGM using MCMC. For each draw, we obtained point and standard error estimates from each of the three study methods (MCMLE, CD, MPLE), evaluating the results with respect to wall-clock estimation time, bias, variance of the estimator, overall accuracy, and calibration (accuracy of estimated standard errors and confidence coverage).  All modeling and analysis was performed using \texttt{statnet} \citep{handcock_statnet:_2008}, specifically using the \texttt{ergm} 4.2.1 \citep{hunter_ergm_2008,krivitsky2022ergm}, \texttt{ergm.count} 4.1.1 \citep{krivitsky_package_2012}, and \texttt{sna} 2.6 \citep{butts_social_2008} libraries.  Our MPLE implementation also made use of the \texttt{Rcpp} library \citep{eddelbuettel2011rcpp}.  The following subsections detail the data and model used, the setup of the estimation procedures, and the performance metrics. %

\subsection{Case study and model definition}
To examine the performance of estimation methods for data with different network sizes and edge value variances, we construct the following study cases. They are based on real-world datasets of migration flows between counties in two U.S. states (New Mexico and North Carolina) \citep{ACSdata}. The New Mexico data consists of 33 nodes and the North Carolina data 100 (henceforth the ``small'' network and the ``large'' network, respectively). For the New Mexico data, we generate two networks with different edge value distributions. The large-variance case uses the count of migrants between each directed county pair as the edge value, which ranges in integers from 0 to 3,862 with standard deviation 201. For comparison, the edge value of the small-variance case takes the natural logarithm of migrant count (plus one), rounded to the nearest integer. Its edge value ranges from 0 to 8, with standard deviation 2. Ideally, we would generate large-variance and small-variance cases for the large network as well. Unfortunately, the large-variance large-network case turned out to be prohibitively computationally expensive for a simulation study comparing all standard methods, and we hence just include the small-variance large network case; the edge value is generated by the same manner discussed above, and the distribution is similar to the small network case, ranging from 0 to 9 with standard deviation 2. Table~\ref{descriptive} displays the descriptive statistics of the three study cases.

\begin{table}[ht]
\caption{Network Descriptive Statistics of the Studied Cases}
\centering
\label{descriptive}
\begin{tabular}{lrrrrrr}
  \hline
  & Network & Binary & \multicolumn{4}{c}{Edge value} \\ \cline{4-7}
 & size & density & min & max & mean & std. dev. \\ 
  \hline
Large-variance small network & 33 & 0.50  & 0 & 3,862 & 46.15 & 200.54 \\ 
  Small-variance small network & 33 & 0.50 &  0 & 8 & 1.64 & 1.95 \\ 
  Small-variance large network & 100 & 0.41 & 0 & 9 & 1.32 & 1.84 \\ 
   \hline
\end{tabular}
\end{table}

Our ground-truth model is created by fitting a count-valued ERGM to the above networks. The sufficient statistics include a Sum term (intercept, the summation of all edge values), a Nonzero term (the count of nonzero edges), three exogenous covariates, and two dependence terms. The exogenous covariates are the population sizes of the sending and receiving counties (called Nodeocov and Nodeicov respectively), and the distance between counties (Edgecov) (all on natural log scale). The first dependence term, mutual, measures the reciprocity of the network, defined as 
\begin{equation*}
g_{\mathrm{m}} =\sum_{(i,j) \in \mathbb{Y} }^{} \min\{y_{ij},y_{ji}\}
\end{equation*}

\noindent The second dependence term, flow, is adapted from a previous model of inter-county migration networks in the United States \citep{huang2022rooted}. It calculates the summation of the volumetric flow of each node, which is the minimum of total inflow and total outflow for a node. It is a count-valued version of two-paths or mixed-two-star terms for binary networks \citep{morris_specification_2008}. Formally,
$$
g_f=\sum_{i \in \mathbb{V}} min \{  \sum_{j \in \mathbb{V} \setminus i} y_{ij} , \sum_{k \in \mathbb{V} \setminus i} y_{ki}      \}
$$

\noindent where $\mathbb{V}$ is the vertex set. This model encompasses a diversity of different sufficient statistics commonly used in valued ERGMs, including graph-level baseline statistics (Sum, Nonzero), covariate effects (Nodeicov, Nodeocov, Edgecov), and dependence terms at the dyadic level (Mutual) and the triadic level (Flow).  Although our aim here is to produce a deliberately simple model for purposes of evaluation (as opposed to a substantively detailed model of migration), our choice of statistics was informed by prior work on migration, and previous empirical analyses on migration-flow networks in particular \citep{boyle_exploring_2014, huang2022rooted,windzio_network_2018,zipf_p1_1946}.

\subsection{Methods under evaluation}

As described above, we estimate parameters from the simulated network draws using three procedures: Contrastive Divergence (CD), Maximum Pseudo-Likelihood Estimation (MPLE), and Monte Carlo Maximum Likelihood Estimation (MCMLE).  Estimation for each method was performed as follows.

For CD, we use the default settings of the \texttt{ergm.count} package, performing 8 Metropolis-Hastings steps, raising one proposal in each step. We also tried using more steps and/or more proposals within each step for CD. As is shown in the Appendix A, its estimation bias and calibration usually do not improve systematically as we increase the tuning parameters, and when it does, it fails to match other methods in comparable time. Therefore, we keep the most time efficient setting in our comparison.

For MPLE, we implement the procedure described in Section~\ref{sec_mple}. To examine the performance of MPLE with various sample sizes, we consider three subsampled MPLE in each study case: the fast, the mid, and the full, which corresponds to uniformly sampling 50\%, 75\% and 100\% of the edge variables in random, respectively; this corresponds to 528, 792, 1056 dyads for small networks, and 4950, 7425, 9900 dyads for the large network. We also consider the impact of multiple cores on execution time, calculating the wall-clock time for models estimated using 1,4, and 20 cores, respectively. In terms of edge support truncation, for small-variance cases, we use a uniform non-edgewise truncation; the support covers integers from 0 to $\lambda$ times the max edge value of the network, where $\lambda=4$ for the small network and $\lambda=1.5$ for the large network as the latter has more information with more edge variables. Edgewise support truncation becomes a powerful tool to reduce computation time for large-variance networks; we set $\lambda=4$ for the doubly-truncated edgewise support (see the second paragraph of Section~\ref{sec_truncation}). We also coerce the support to include integers from 0-value to the 80\% quantile of the edge distribution for every edge variable, whose upper bound ranges from 43 to 48 and is typically close to the mean of the edge distribution. This is a conservative scheme for an edgewise support truncation with wide intervals and without coarsening, but is fast enough to have one order of magnitude less wall-clock time than MCMLE.

For MCMLE, we use the stochastic approximation method in \texttt{ergm.count}, a workhorse method that is also implemented and served as the default method in PNET (\citealt{wang_pnet_2009}). We made a few adjustments from the default setting to improve its performance based on the data structure of the study case and our exploratory experiments. First, we set the proposal distribution of Metropolis-Hastings algorithm in MCMC to be random, where every dyad has equal chance to be toggled. By default, the proposal distribution in \texttt{statnet} favors toggling nonzero edges than nulls, with the rationale that social networks tend to be sparse (for binary networks); we revoke this penalty towards nulls since the binary density of the network is high (0.4-0.5) in the study cases (removing the need for biased proposals), and the random proposal reduces the computational time of the MCMLE. Second, by default, the MCMC thinning interval is 1024 and the sample size of network statistics in each distribution returned by the algorithm is also 1024.\footnote{We follow the default of \texttt{statnet} that sets MCMC burnin as 16 times the length of the thinning interval.} This setting is sufficient for the small-variance small network case, but for the other two cases, we increase these two parameters to be ten thousand; this helps with convergence and further increasing those parameters no longer brings performance gain based on our experiments. We evaluate the use of both CD and MPLE as seeding methods for MCMLE, referring to them as CD-MCMLE and MPLE-MCMLE, respectively. We employ default settings for CD, given that longer chains do not consistently enhance performance (see Appendix A). For the small-variance small network case, we use the ``mid'' MPLE for MCMLE seeding, but for the large-variance and the large network cases, we use the ``fast'' MPLE since the fast MPLE yields good-enough point estimations. It turns out that CD-MCMLE fails to converge for some of the large-variance cases; we detect this by examining the MCMC diagnostics plot, and reruning those cases until convergence (details in Table~\ref{tab:cdmcmle} in Section~\ref{sec_results}).  We also turn off the bridge sampler that calculates the log likelihood to reduce the MCMLE computational time, since this is not involved in estimation.

All models were fit on a 44-core server, with 256GB RAM. The processors are dual Intel Xeon E5-2699 2.2GHz CPUs (22 cores/CPU). Estimation using R 4.1.1 was performed on Ubuntu 20.04.1. All procedures reported are based on a single core, except for the multi-core MPLE conditions. 

\subsection{Evaluation criteria}

Since the methods of interest involve speed/quality trade-offs, it is necessary to evaluate these two dimensions simultaneously. To evaluate computational cost, we compute the wall-clock time for each method, as mean seconds needed to fit the target model to a simulated network. Since the speed of MPLE is dependent upon the number of parallel processes, we repeat the process using 1, 4, and 20 cores, respectively.

To evaluate estimation quality, we consider the bias, variance, overall accuracy, and calibration of each estimator using the following metrics.

We first compute the absolute relative bias (ARB) of estimators for each coefficient, using 
\begin{equation*}
ARB=\left|\frac{1}{m}\sum_{i=1}^{m}{ \frac{\hat{\theta_i}-\theta}{\theta} } \right|
\end{equation*}

\noindent taking the average across $m$ experiments, where $\hat{\theta}$ is the estimator and $\theta$ is the true value from the model that simulated the networks. The smaller the ARB, the less bias is introduced by the estimation procedure.

We also compute the variability for each estimator, via the (true) standard error of the estimated coefficient. Formally,
\begin{equation*}
SE=\sqrt{\frac{1}{m}\sum_{i=1}^{m}{(\hat{\theta_i}-\frac{1}{m}\sum_{j=1}^{m}{\hat{\theta_j})}^2}}
\end{equation*}

\noindent The smaller the variability, the more efficient and more precise the estimator is.

While bias and variance are each important, we are also interested in the total accuracy of the estimator (the extent to which it deviates, on average, from the true value).  We measure this via the root-mean-square error (RMSE) i.e. 
\begin{equation*}
RMSE=\sqrt{\frac{1}{m}\sum_{i=1}^{m}{(\hat{\theta_i}-\theta)^2}}    
\end{equation*}

\noindent The smaller the RMSE, the more accurate the estimator is on average.

Finally, we consider how well calibrated each estimator is, in terms of the associated estimates of uncertainty.  To evaluate the bias in our second moment estimate, we compare the real standard error $se$ and the estimated standard error $\hat{se}$ using 

\begin{equation*}
Calibration=\log\left[\frac{1}{m}\sum_{i=1}^{m}{\frac{\hat{se_i}}{se}}\right]
\end{equation*}

\noindent A positive number suggests the method is conservative, while a negative number suggests the method is overconfident.  We also examine confidence coverage, specifically the proportion of cases in which the nominal 95\% confidence interval (CI) for each parameter actually covers the true coefficient.  Specifically, the coverage rate is computed by 

\begin{equation*}
Coverage=\frac{1}{m}\sum_{i=1}^{m}{\mathbbm{1}\{\theta \in [\hat{\theta_i}-z \cdot \hat{se_i},\hat{\theta_i}+z \cdot \hat{se_i} ]}\}
\end{equation*}

\noindent where $z=1.96$ for 95\% CI.  The closer to 95\% the coverage is, the better calibrated the estimate is. Coverage rates above 95\% suggest that the method is conservative, while coverage rates less than 95\% suggest that the method is overconfident.

\section{Results} \label{sec_results}
\subsection{The small-variance, small network case}
\begin{figure}[p]
\centering
    \begin{subfigure}{\textwidth}
        \includegraphics[width=\textwidth]{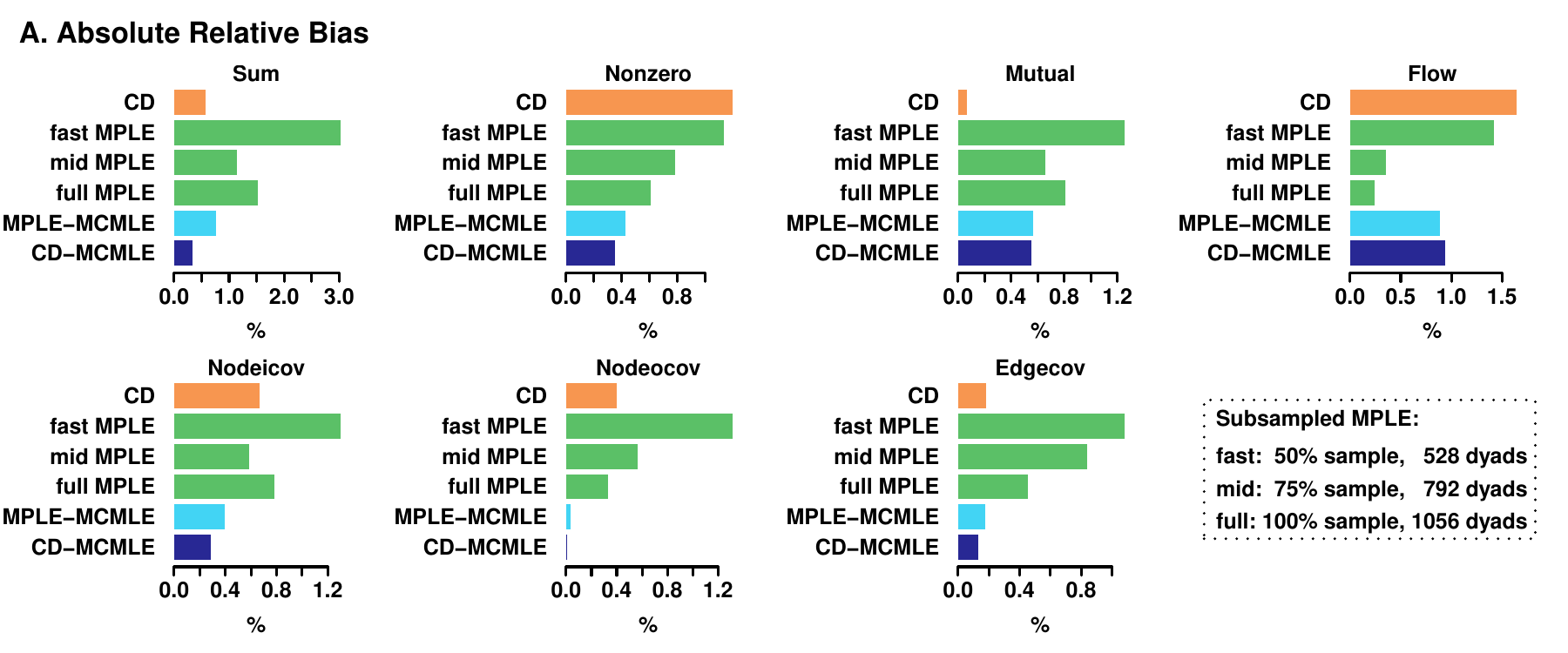}
    \end{subfigure}
    \begin{subfigure}{\textwidth}
        \includegraphics[width=\textwidth]{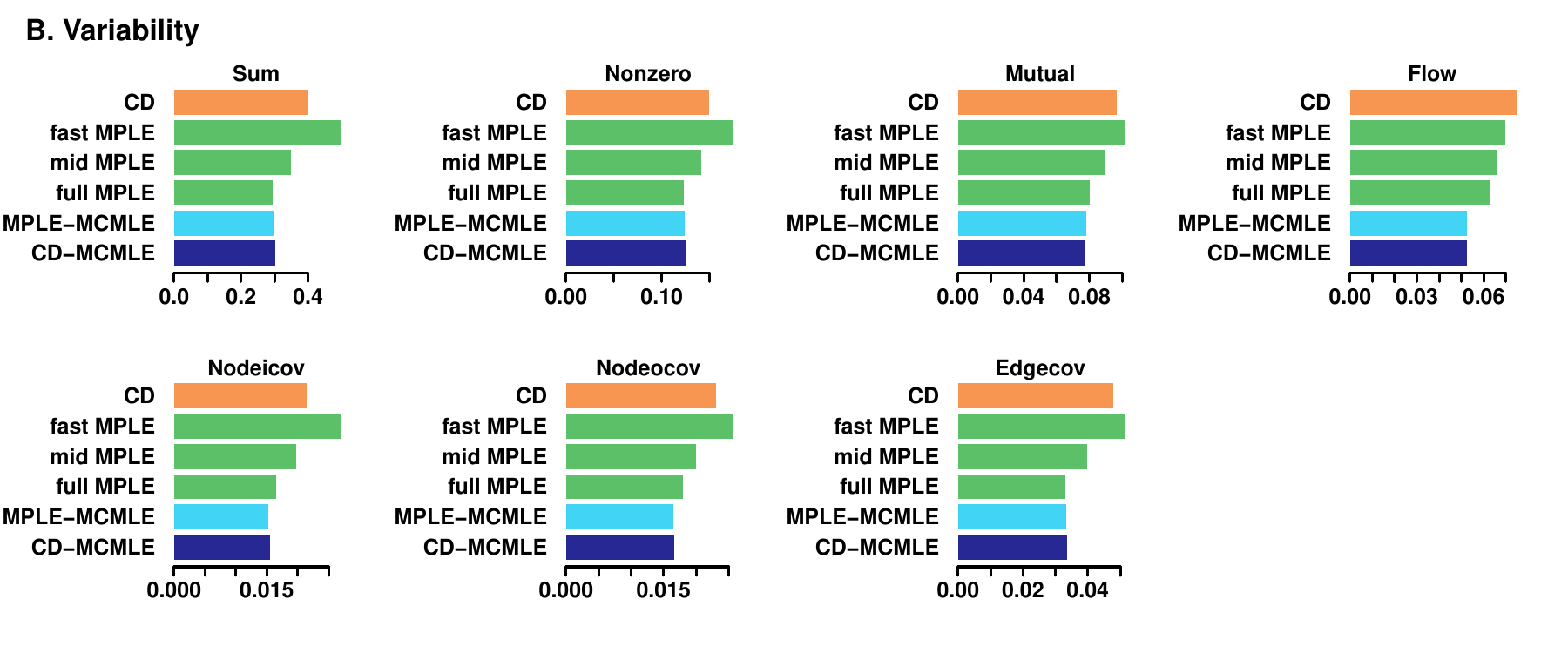}
    \end{subfigure}
    \begin{subfigure}{\textwidth}
        \includegraphics[width=\textwidth]{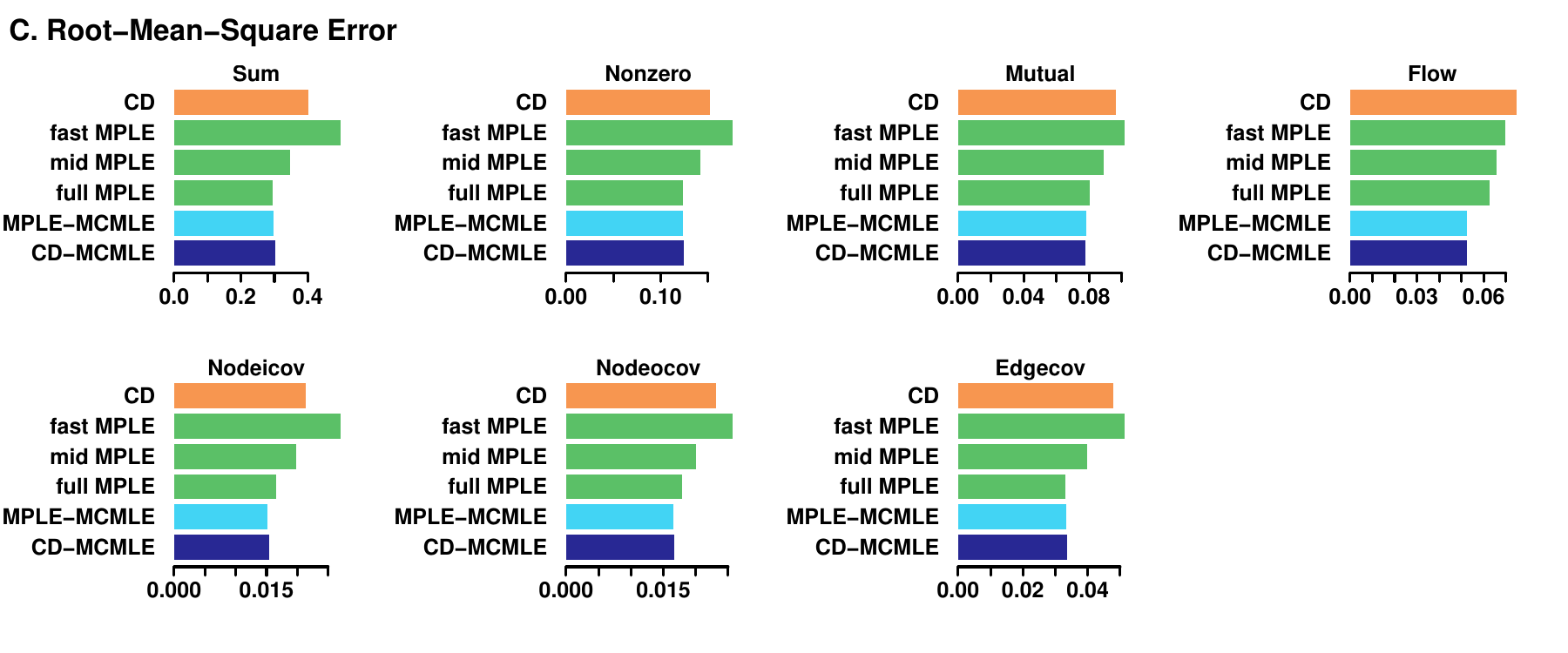}
    \end{subfigure}
    \caption{\leftline{Bias, variability, and RMSE of small-variance small network}}
\label{fig:easy1}
\end{figure}

\begin{figure}[p]
\centering
    \begin{subfigure}{\textwidth}
        \includegraphics[width=\textwidth]{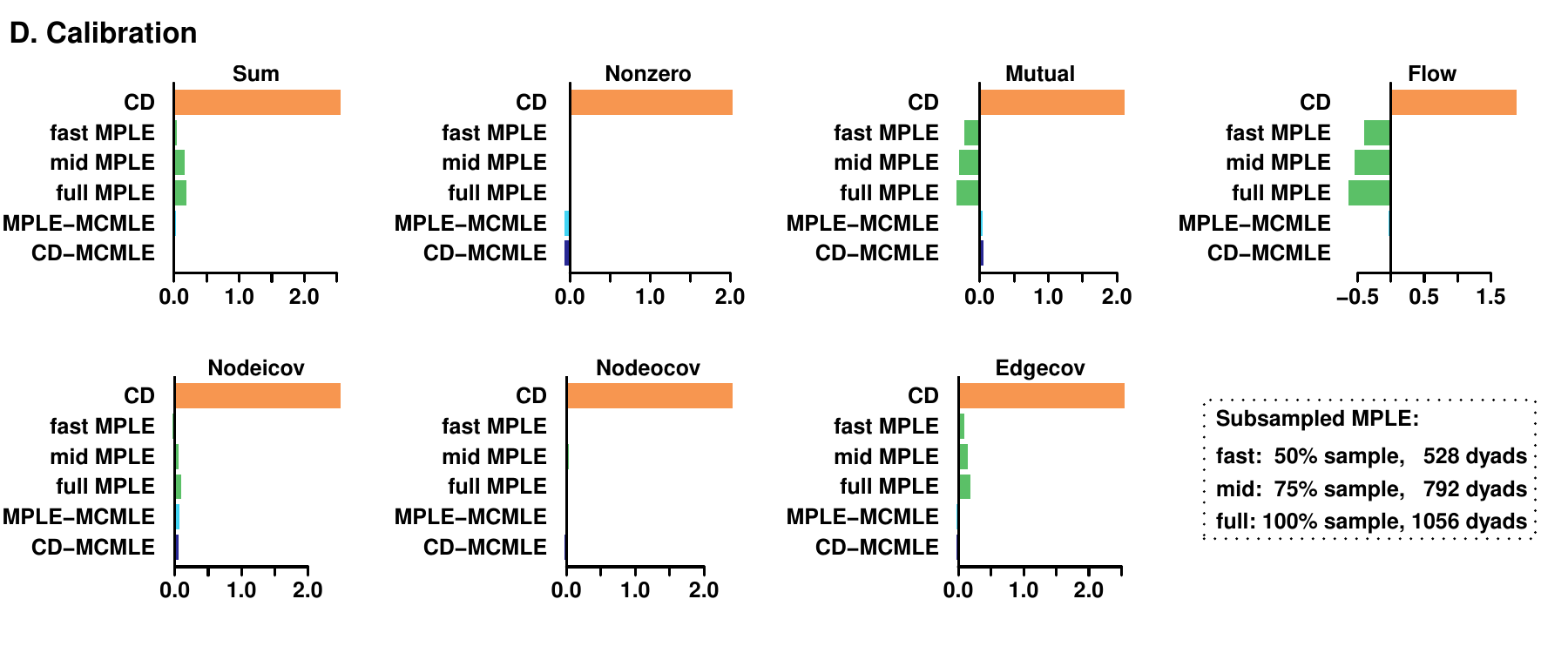}
    \end{subfigure}
    \begin{subfigure}{\textwidth}
        \includegraphics[width=\textwidth]{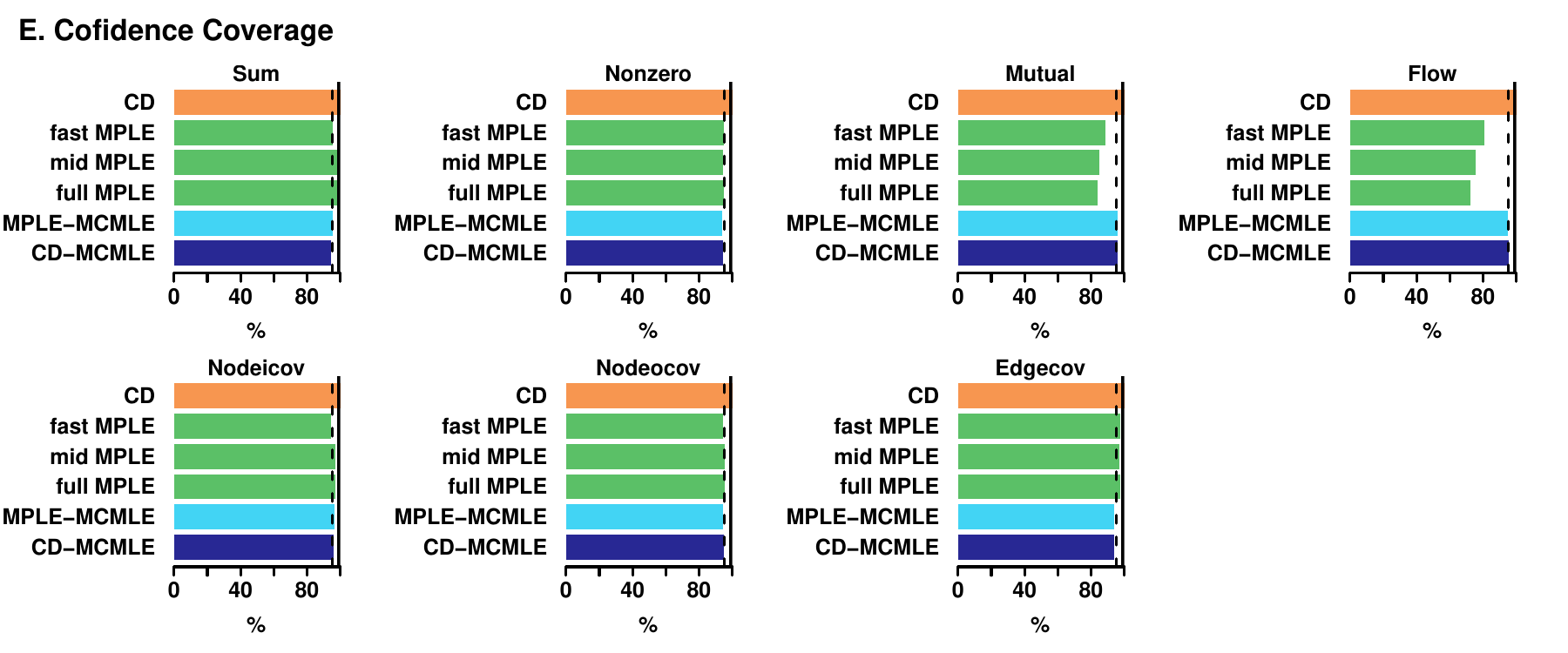}
    \end{subfigure}
    \begin{subfigure}{\textwidth}
        \includegraphics[width=\textwidth]{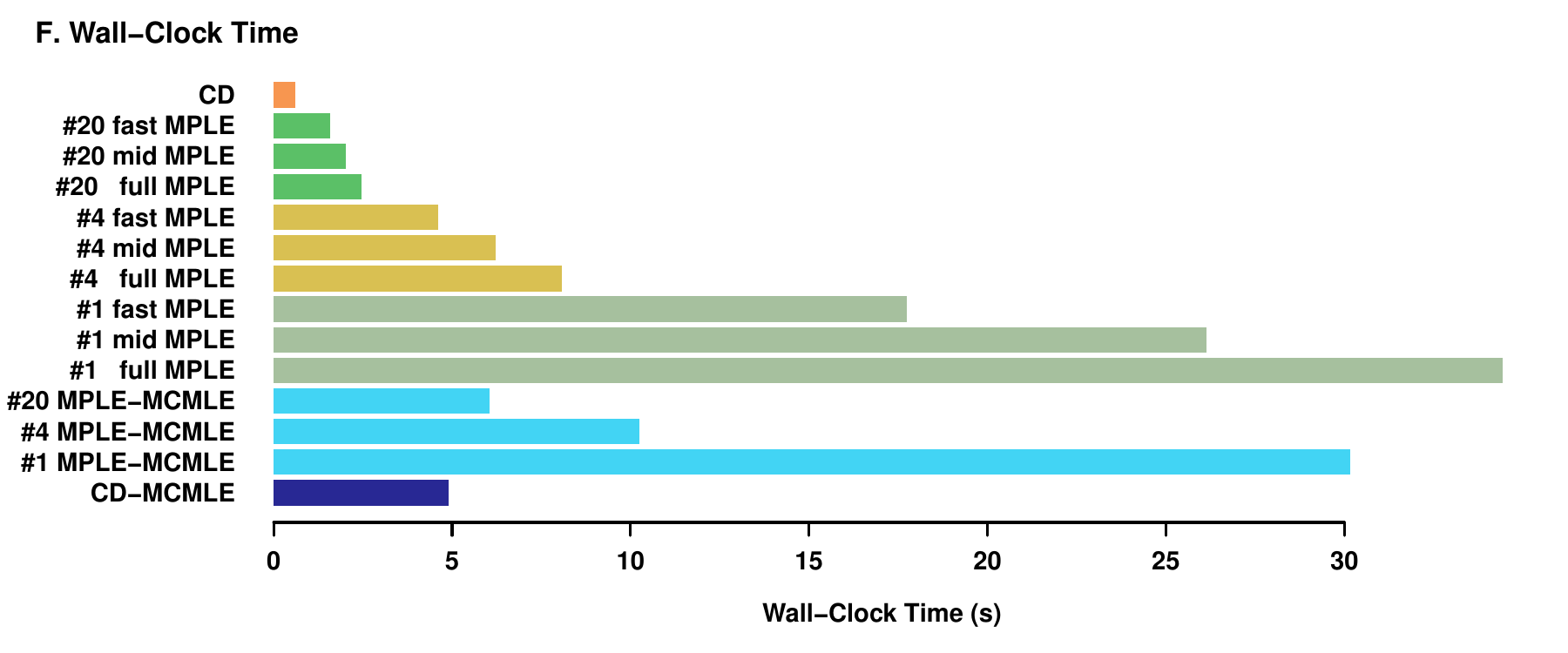}
    \end{subfigure}
    \caption{\leftline{Calibration, confidence coverage, and wall-clock time of small-variance small network}}
\label{fig:easy2}
\end{figure}

Starting from the simplest case of the small-variance small network, we display the performance of each method in Figures~\ref{fig:easy1} and~\ref{fig:easy2}. Panel A in Figure~\ref{fig:easy1} shows the absolute relative bias (ARB) of the coefficient estimates. It shows that all methods produce very small numerical biases, 3\% or less across all covariates and methods. CD and fast MPLE introduces larger biases; but as the sample size of MPLE increases, its bias reduces and gets close to that of MCMLE, seeded by either CD or MPLE. This finding is consistent with research on binary ERGMs finding that the MPLE introduces little bias in parameter estimation \citep{van_duijn_framework_2009,schmid_exponential_2017}. The lack of appreciable bias is an encouraging sign, suggesting that point estimation of valued ERGMs for small-variance small network is easy to acquire using whichever method we tested.%

We also evaluate the (im)precision or variability of estimation (sometimes called efficiency), i.e. the true standard error of each estimator. Panel B in Figure~\ref{fig:easy1} shows that the variability of the MPLE decreases with more edges sampled, and that full MPLE and MCMLE are the most efficient methods. In general, variations of estimators using all methods are close to each other, suggesting that they have similar efficiency.

We then evaluate the total accuracy of estimators using the root-mean-square error (RMSE). A more holistic metric, the accuracy measurement combines bias and variation of estimation evaluated above, and smaller RMSE is preferred. Panel C of Figure~\ref{fig:easy1} displays RMSE scores, whose distribution is almost identical to the variability scores in Panel B. The similarity between RMSE and variability reveals that biases contribute very little to the total RMSE, with accuracy being dominated by the performance in variability. Methods with good variability score thus also have decent accuracy. Full MPLE and MCMLE has the smallest RMSE, though RMSEs for all methods under evaluation show only mild differences.

Besides performance in coefficient estimation, performance in estimating uncertainties is also evaluated, shown in the first two panels in Figure~\ref{fig:easy2}. Panel D displays calibration of each method. An indicator of the bias in standard error estimation, a positive calibration score suggests overestimation of the uncertainty, and a negative calibration suggests underestimation. Noticeably, CD overestimates standard errors for all covariates by a large margin; CD's calibration scores are all above 1.9, indicating that the estimated standard error is more than 6.6 (i.e., $e^{1.9}$) times its true value. We experimented with different tuning parameters for CD, but could not find settings with both improved calibration and reasonable execution time (see Table~\ref{tab:cdcalib} in Appendix A). In summary, CD is too conservative to offer useful uncertainty estimations of covariates for small-variance small networks.

For MPLE, we find that it underestimates the uncertainties for the dependence terms (mutual and flow), but overestimates uncertainties for non-dependence terms, though the degree of inflation is very small. Previous simulation studies found similar patterns for MPLE on binary ERGMs \citep{lubbers_comparison_2007,van_duijn_framework_2009}. Interestingly, the bias in standard error estimation increases with the sample size for MPLE. This is in part because the bias of estimation in statistical uncertainty is trivial when the numerical uncertainty is the main source of uncertainty for the MPLE (as is the case with small sample sizes); as the numerical uncertainties decrease with more edges sampled, the bias in statistical uncertainty becomes non-negligible. The calibration of MPLE is not as far off as that of CD, but its underestimation of standard errors is noticeable for dependence terms in the small-variance small network case.

While CD and the MPLE show varying degrees of error in standard deviation, both CD-MCMLE and MPLE-MCMLE have almost perfect calibration, suggesting that MCMLE is the best method for standard error estimation of small-variance small network.

Another metric that considers uncertainty estimation, confidence coverage is the proportion of model fits in which the true value of a given coefficient is covered by the estimated 95\% confidence interval (CI), as is shown in Panel E of Figure~\ref{fig:easy2}, where the dotted line is the 95\% reference line and the solid line represents 100\%. The figure tells a similar story to Panel D's calibration score, because confidence coverage performance is largely determined by performance in calibration of uncertainty when the bias of coefficient estimations is small. The figures show that CD overestimates standard errors so much that its CIs always cover the true value, making them conservative but uninformative. The MPLE's CIs cover the true values more than 95\% for non-dependence terms, but under-cover the true values for the dependence term, with this deviation becoming larger as sample size increases. On the other hand, both CD-MCMLE and MPLE-MCMLE have coverage rates that are extremely close to 95\%, showing its characteristic calibration advantage for small-variance small network of valued ERGMs estimation. %

Lastly, Panel F of Figure~\ref{fig:easy2} displays the wall-clock time of each method. As expected, the wall-clock time for computing the MPLE can be greatly compressed by using a sample of edges to approximate the joint pseudo-likelihood function, or by using multiple cores to calculate conditional likelihoods. The fastest methods are CD followed by MPLE using 20 processors. Overall, the wall-clock time for different methods is short and varies modestly for this simple computation case, costing half a minute at most. Interestingly, while MCMLE is commonly believed to be a slow method, it is very fast in this simple case.

In summary, for small-variance small network data, all methods offers accurate and minimally-biased point estimates. CD offers uninformatively conservative uncertainty estimates, and MPLE's uncertainty estimation for dependence terms is noticeably overconfident. All methods are reasonably fast in this regime. With great performance in all metrics, MCMLE is an ideal method for valued ERGMs estimation for small-variance small network data.

\subsection{The small-variance, large network case} \label{sec_NC}
The small-variance large network case has similar edge value distribution to the previous small network case, but its network size is 3 times bigger, meaning that its dyad count is 9.4 times the count of the previous case. Comparing these two cases offers insights about the influence of network size on estimation performance for each method.

\begin{figure}[p]
\centering
    \begin{subfigure}{\textwidth}
        \includegraphics[width=\textwidth]{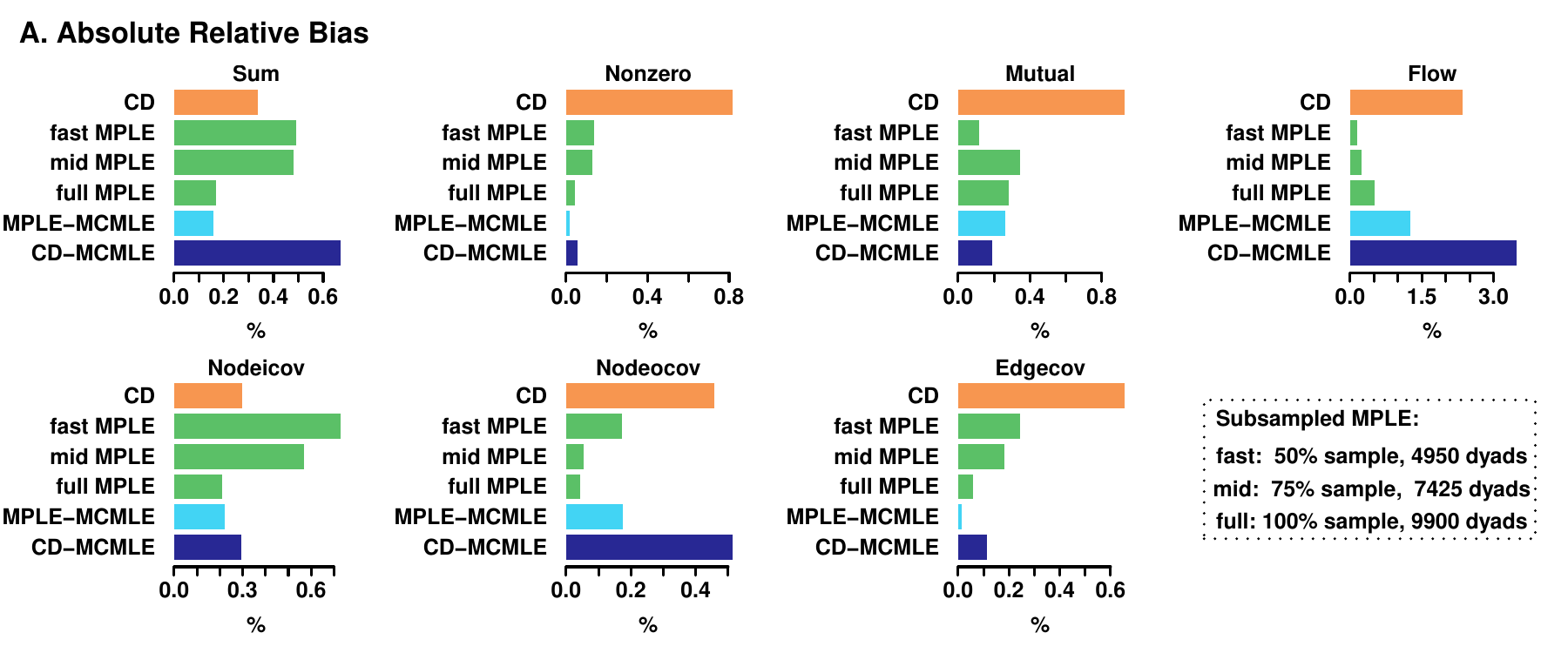}
    \end{subfigure}
    \begin{subfigure}{\textwidth}
        \includegraphics[width=\textwidth]{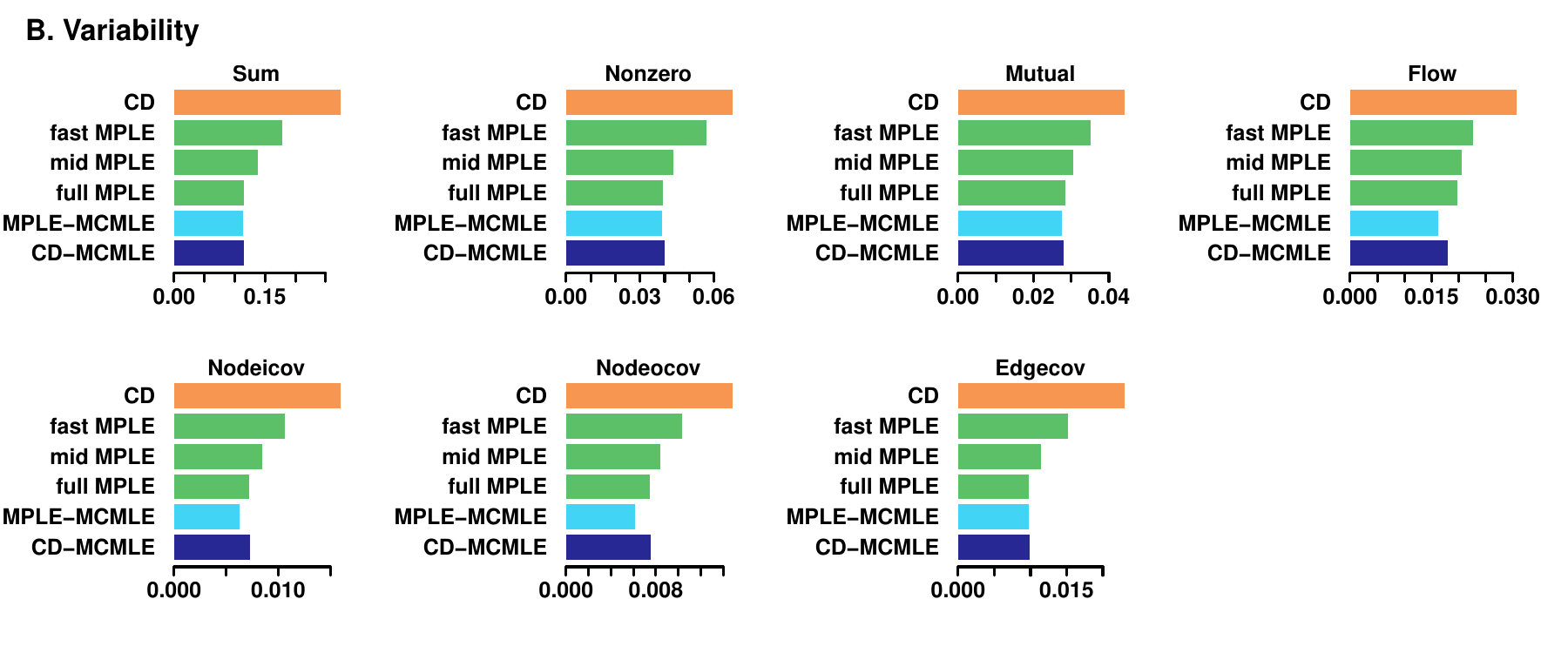}
    \end{subfigure}
    \begin{subfigure}{\textwidth}
        \includegraphics[width=\textwidth]{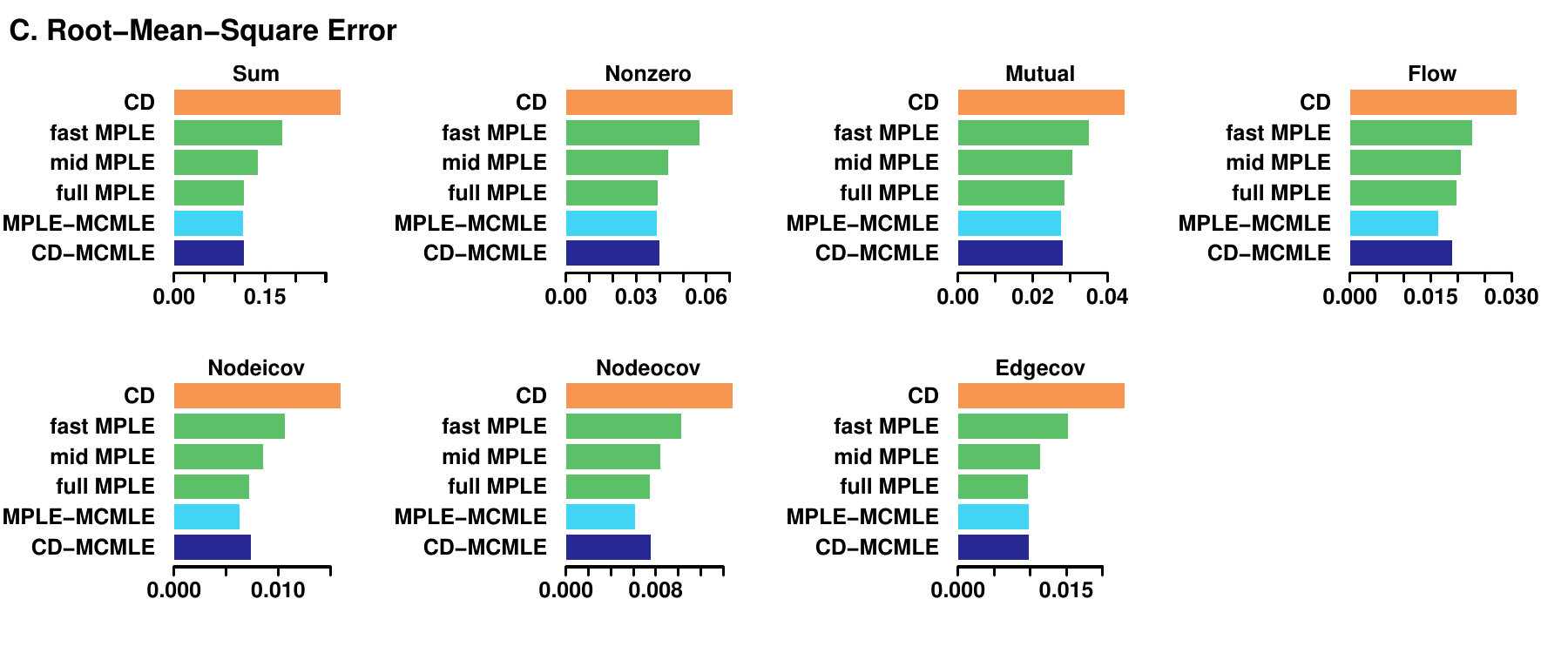}
    \end{subfigure}
    \caption{\leftline{Bias, variability, and RMSE for small-variance large network}}
\label{fig:nc1}
\end{figure}

\begin{figure}[p]
\centering
    \begin{subfigure}{\textwidth}
        \includegraphics[width=\textwidth]{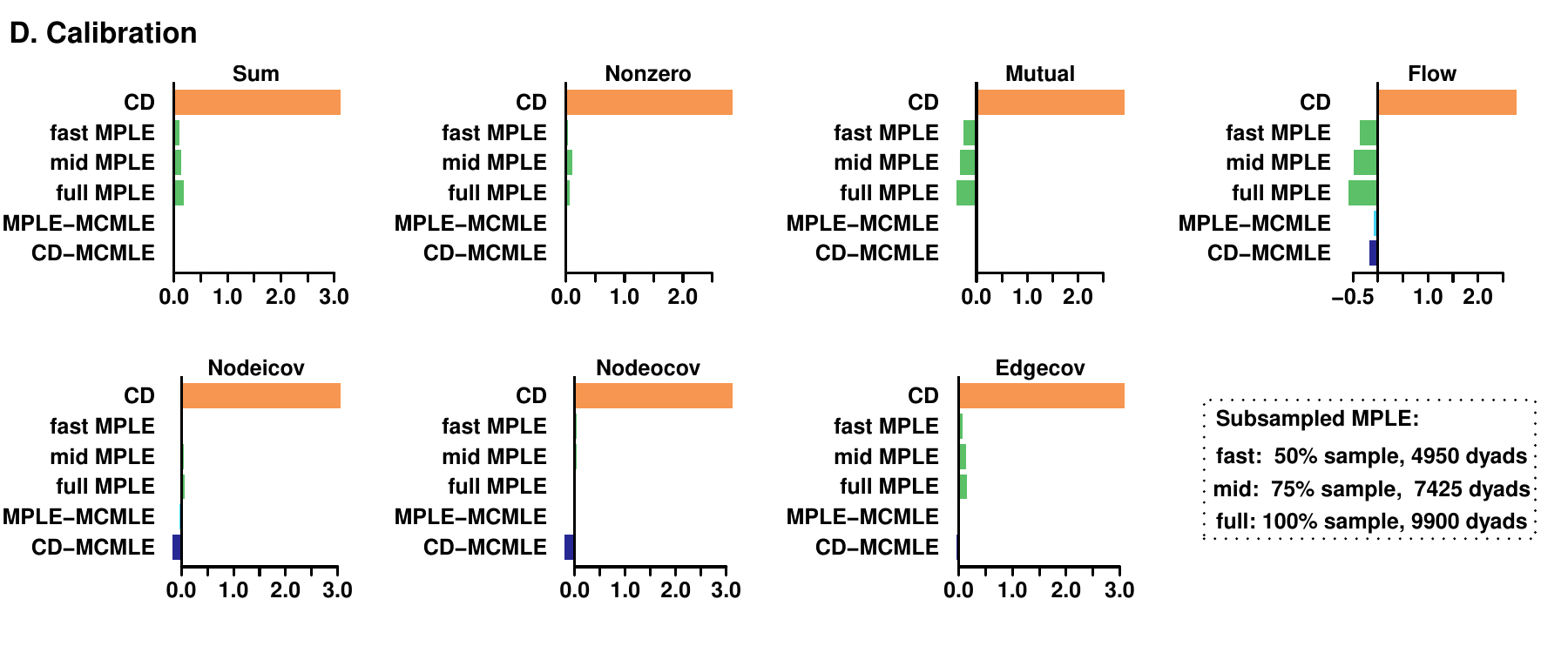}
    \end{subfigure}
    \begin{subfigure}{\textwidth}
        \includegraphics[width=\textwidth]{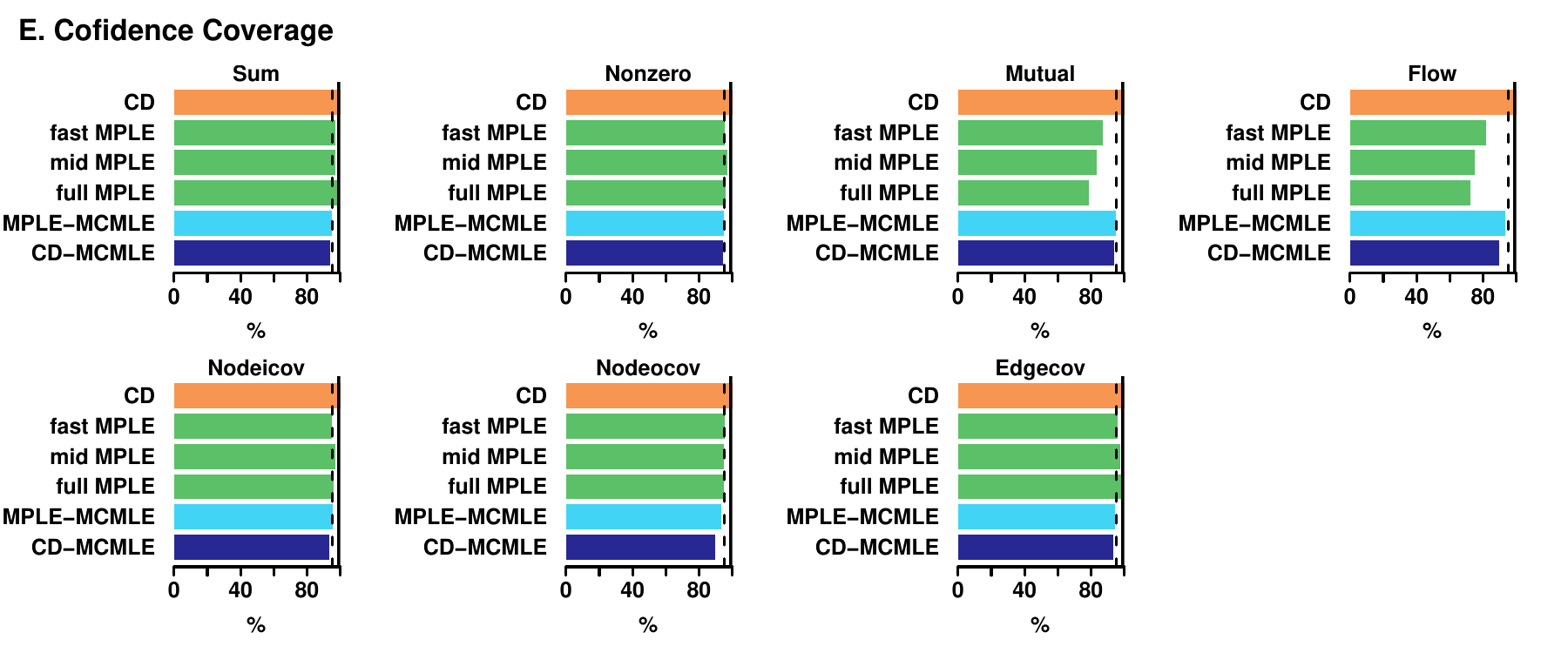}
    \end{subfigure}
    \begin{subfigure}{\textwidth}
        \includegraphics[width=\textwidth]{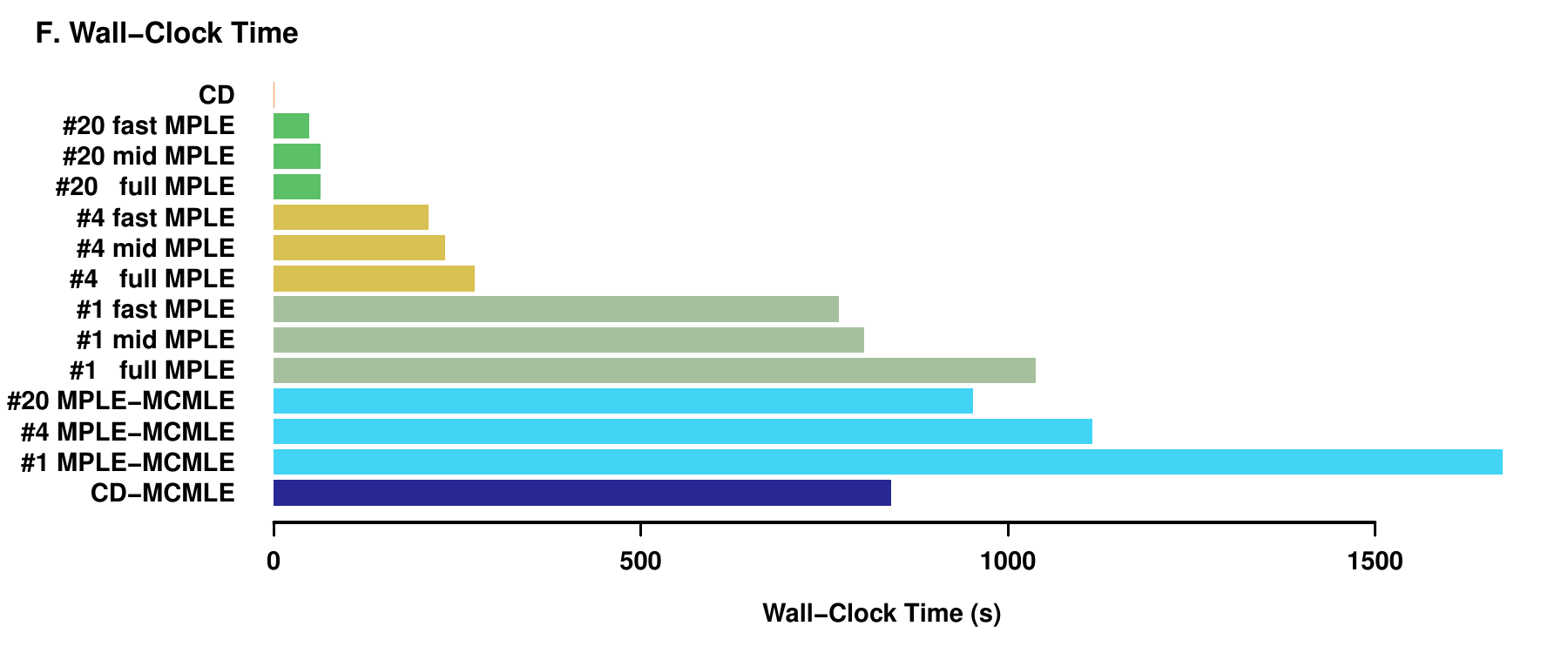}
    \end{subfigure}
    \caption{\leftline{Calibration, confidence coverage, and wall-clock time for small-variance large network}} %
\label{fig:nc2}
\end{figure}

Figure~\ref{fig:nc1} shows that, again, all methods introduce very little bias. One noticeable difference is that the bias of subsampled MPLE gets smaller in comparison with other methods. This suggests that the absolute number of edge variables sampled influences the performance of subsampled MPLE; for small networks, one needs to sample a larger proportion of edges, while for large networks, the percentage can be lower. This means that for the large network case, the fastest MPLE is already less biased than CD (whose bias from a larger tuning parameter setting gets outperformed by following up CD with MCMLE, see Table~\ref{tab:cdbias} in Appendix A); this translates to the less biased performance of MPLE-seeded MCMLE, compared to the CD-seeded MCMLE as the figure shows. Panels B and C in Figure~\ref{fig:nc1} reveal almost identical patterns in variability and RMSE compared to the small-variance small network case. Variability decreases as MPLE's sample size increases, and all methods share similar variability. The RMSE distributions resembles those of the variability as the bias of all methods are largely ignore-able.

Comparing Panels D and E in Figure~\ref{fig:nc2} with those in the previous Figure~\ref{fig:easy2} suggests that patterns of uncertainty estimation performance across  methods are generally invariant to network size. We again find that CD greatly overestimates uncertainty, leading to over-coverage of the confidence intervals; MPLE underestimates the uncertainty for dependence terms and the confidence intervals under-covers their true values. MCMLE offers great uncertainty estimation again, although, on close inspection, CD-MCMLE shows a (very) small tendency towards overconfidence that the MPLE-MCMLE lacks.

Panel F in Figure~\ref{fig:nc2} shows that the wall-clock time of MPLE and MCMLE scales with the network size, while that does not apply to CD. Subsampling and use of multiple cores effectively reduce the computational time of MPLE; this advantage becomes larger with network size, since larger networks require a greater share of computing time to be used for change score calculation.  Since changescore calculations are embarrassingly parallelizable, gains from multi-core calculations grow accordingly in this regime.  Overall, this scenario shows clear superiority of CD for computational time, followed by MPLE using subsampling and multi-core strategies.  MCMCLE becomes quite slow here (with mean times between $\approx 13$ minutes and roughly half an hour), making speed a potentially important consideration.

To recap, comparing the two small-variance cases with small and large network sizes suggests that the estimation quality of each method is largely invariant to the network size. The larger number of edge variables for the large network means that subsampled MPLE requires a smaller share of edge variables for good performance, and fast MPLE becomes a less biased and a better seeding method than CD for MCMLE (though the difference is small in the study cases). All methods have good first-order performance, though MCMLE is clearly superior for calibration (with CD being unacceptably poor).  The major performance difference coming from the network size is that the superiority in computational efficiency for CD and multi-core subsampled MPLE becomes substantial when the network size grows.

\subsection{The large-variance, small network case}

Given the consistency seen in the two small-variance cases, we might expect the large variance, small network case to behave similarly.  However, we observe very different results in the small network case when the variance of edge values becomes large.  First of all, we observe that CD-MCMLE simply fails to converge for some of the simulated networks, as reflected by their MCMC diagnostics plots; here we follow the common procedure of rerunning them until convergence, though failure to attend to diagnostics could lead to problems in casual use. Table~\ref{tab:cdmcmle} summarizes the number of rounds CD-MCMLE went through before seeing convergence.  Overall, it took 1.22 rounds on average for CD-MCMLE to converge (with a few cases taking more than five). The following results are based on their final (converged) rounds, as estimators from the failed rounds were very far from the true values.

\begin{table}[ht]
\caption{\leftline{Number of rounds for CD-MCMLE before convergence}}
\label{tab:cdmcmle}
\begin{tabular}{rrrrrrrr}
  \hline
N of Rounds & 1 & 2 & 3 & 4 & 5 & 6 & 7 \\ 
  \hline
Count & 434 & 38 & 17 & 9 & 0 & 1 & 1 \\ 
Percentage (\%) & 86.8 & 7.6 & 3.4 & 1.8 & 0 & 0.2 & 0.2 \\  
   \hline
\end{tabular}
\end{table}

\begin{figure}[p]
\centering
    \begin{subfigure}{\textwidth}
        \includegraphics[width=\textwidth]{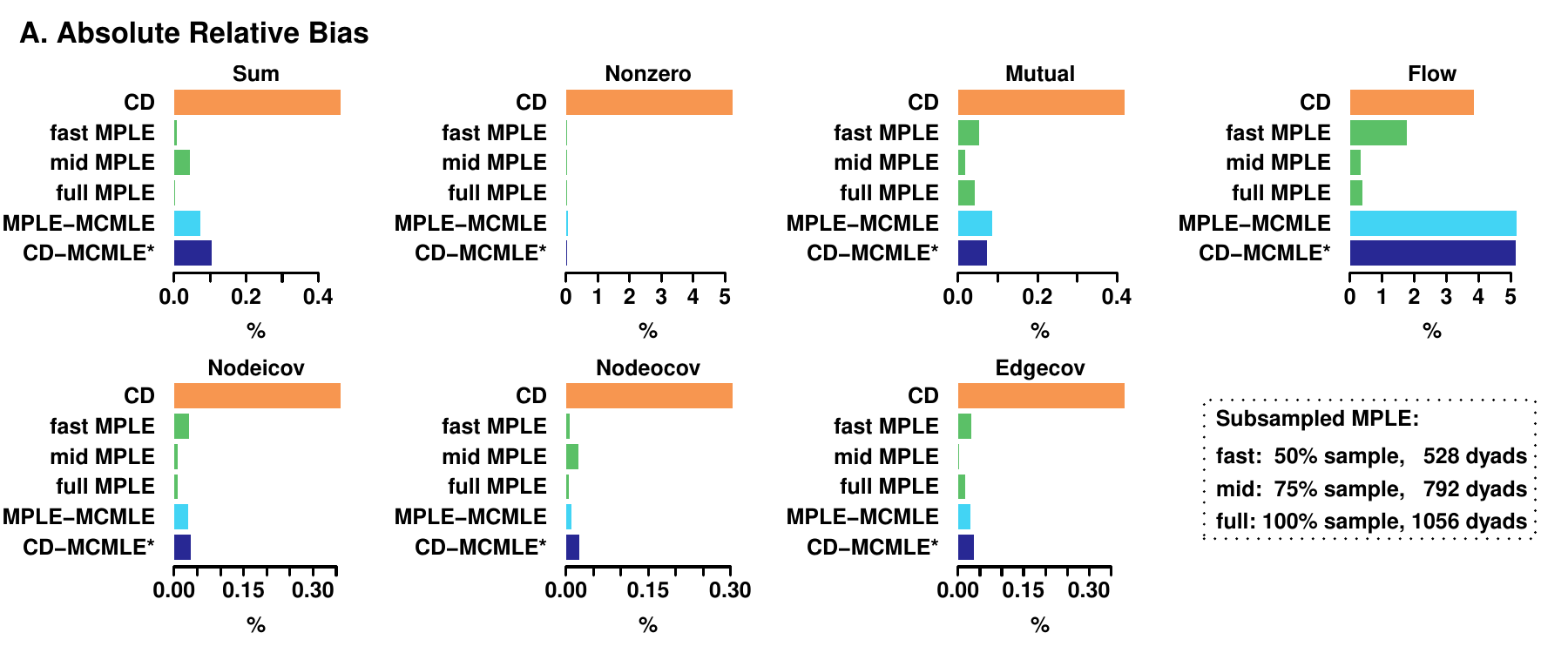}
    \end{subfigure}
    \begin{subfigure}{\textwidth}
        \includegraphics[width=\textwidth]{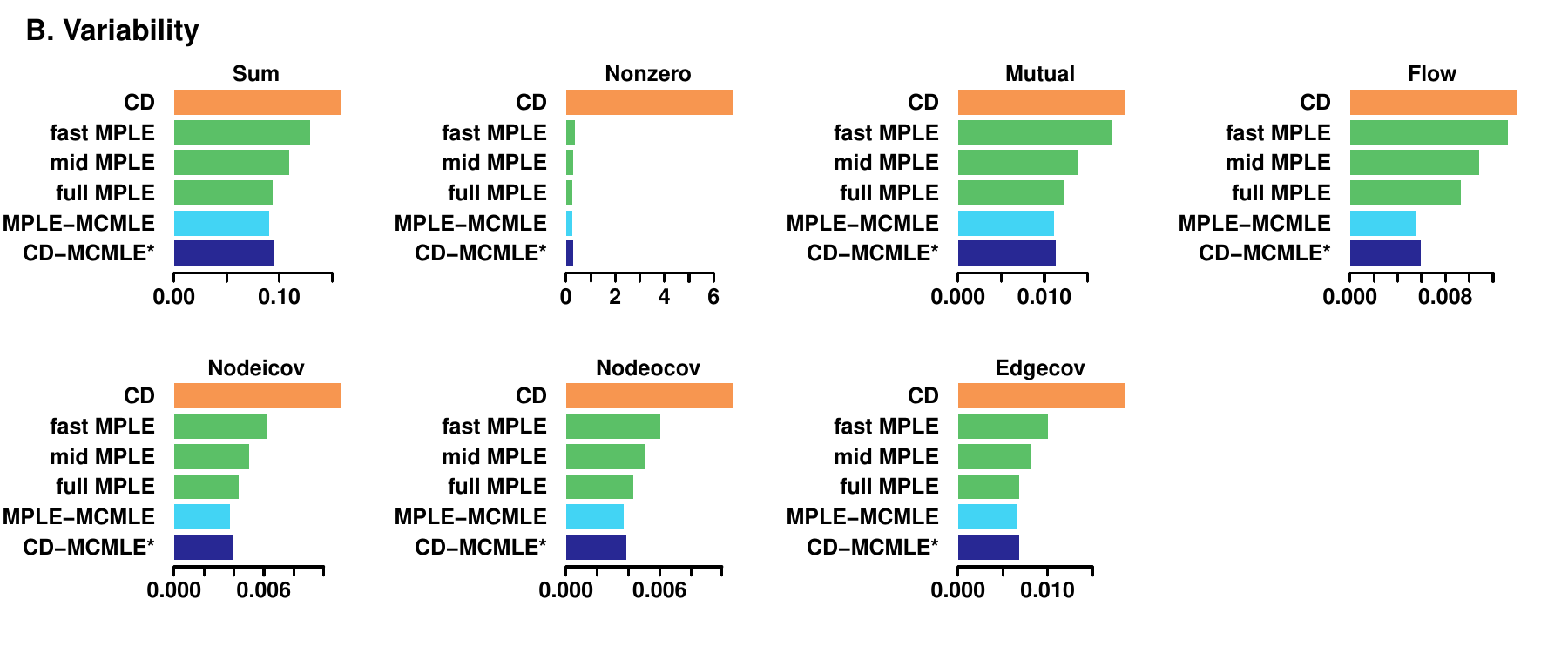}
    \end{subfigure}
    \begin{subfigure}{\textwidth}
        \includegraphics[width=\textwidth]{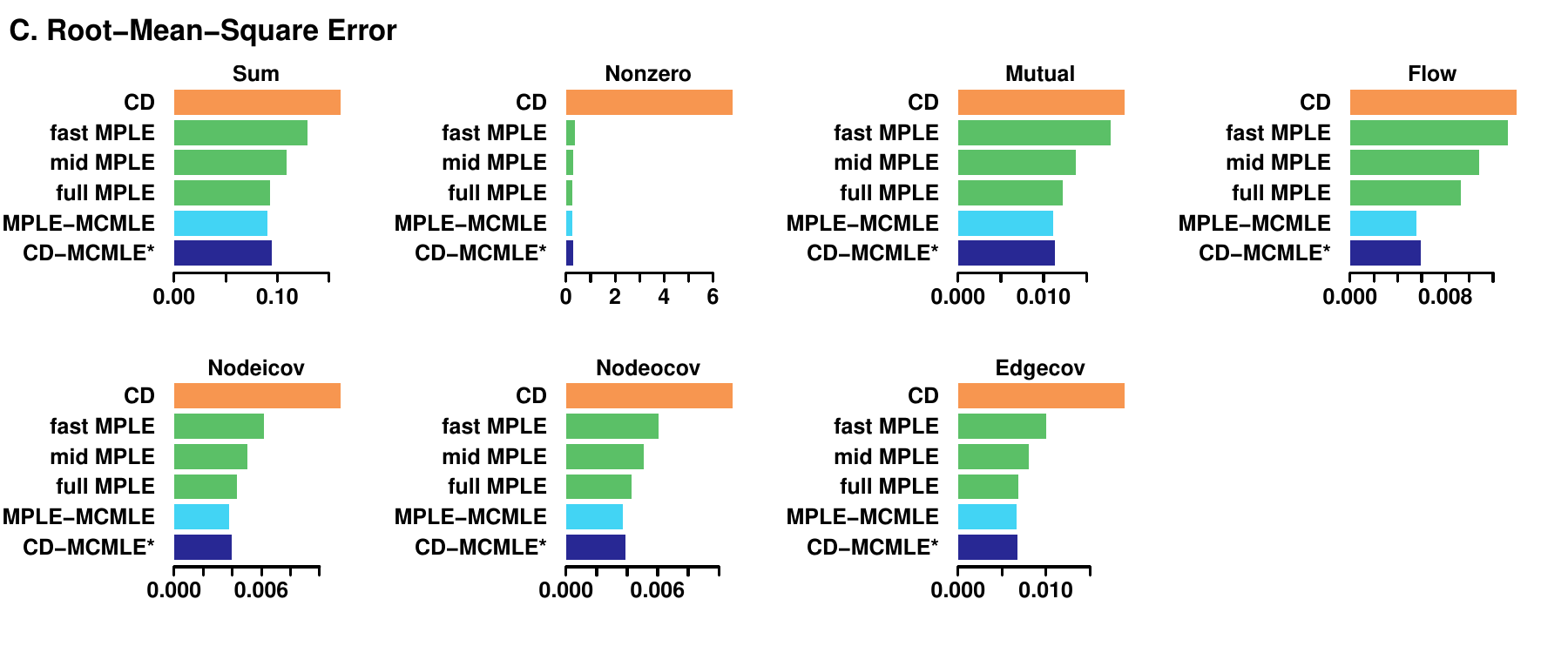}
    \end{subfigure}
    \caption{\leftline{Bias, variability, and RMSE of large-variance small network}}
\label{fig:hard1}
\caption*{\leftline{\emph{Note:} CD-MCMLE* results are from their final rounds with convergence.}}
\end{figure}

Panel A in  Figure~\ref{fig:hard1} shows that CD introduces relatively larger biases for the large-variance case, 5.2\% and 3.8\% for the nonzero and the flow terms, respectively. Although those biases are arguably not huge, they lead to failures in convergence for CD-MCMLE. On the contrary, MPLE introduces very little bias in its estimates, the largest bias of 1.8\% coming from the fast MPLE for the flow term. That makes it an excellent seeding method, and indeed all MPLE-MCMLE models converged in their first attempt. This signifies that,  similar to the binary ERGM scenario, MCMLE for valued ERGMs is sensitive to the seeding quality, and small improvements in biases of the seeding methods can make a difference. Another notable feature is that MPLE actually outperforms MCMLE in the bias metric, with the latter having a bias of about 5\% for the triadic dependence term (flow).

Panel B in Figure~\ref{fig:hard1} shows that CD generally has larger variability than other methods, especially for the nonzero term. MPLE's variability decreases with more edge variables sampled, and gets close to MCMLE when all edges are utilized. Panel C in Figure~\ref{fig:hard1} demonstrates that, again, when biases are generally small, the accuracy metric resembles that of the variability. 

\begin{figure}[p]
\centering
    \begin{subfigure}{\textwidth}
        \includegraphics[width=\textwidth]{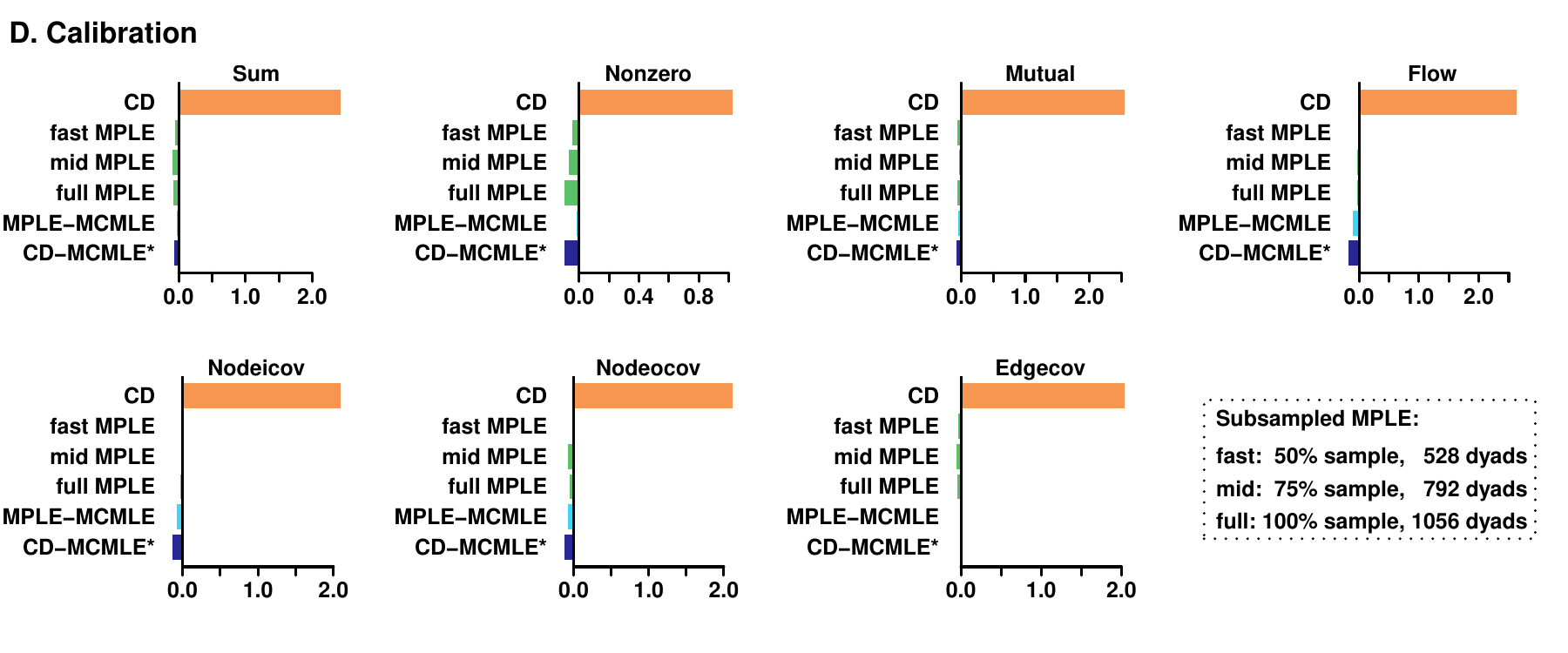}
    \end{subfigure}
    \begin{subfigure}{\textwidth}
        \includegraphics[width=\textwidth]{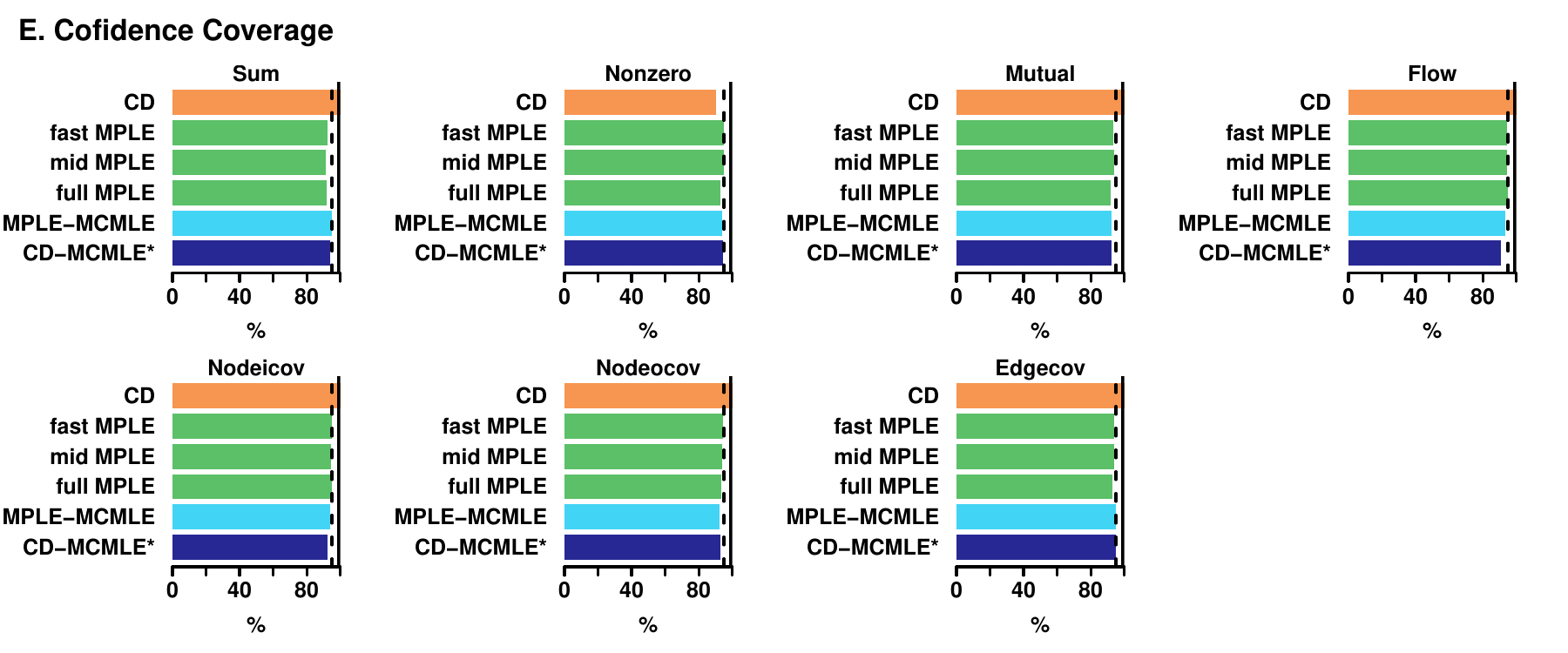}
    \end{subfigure}
    \begin{subfigure}{\textwidth}
        \includegraphics[width=\textwidth]{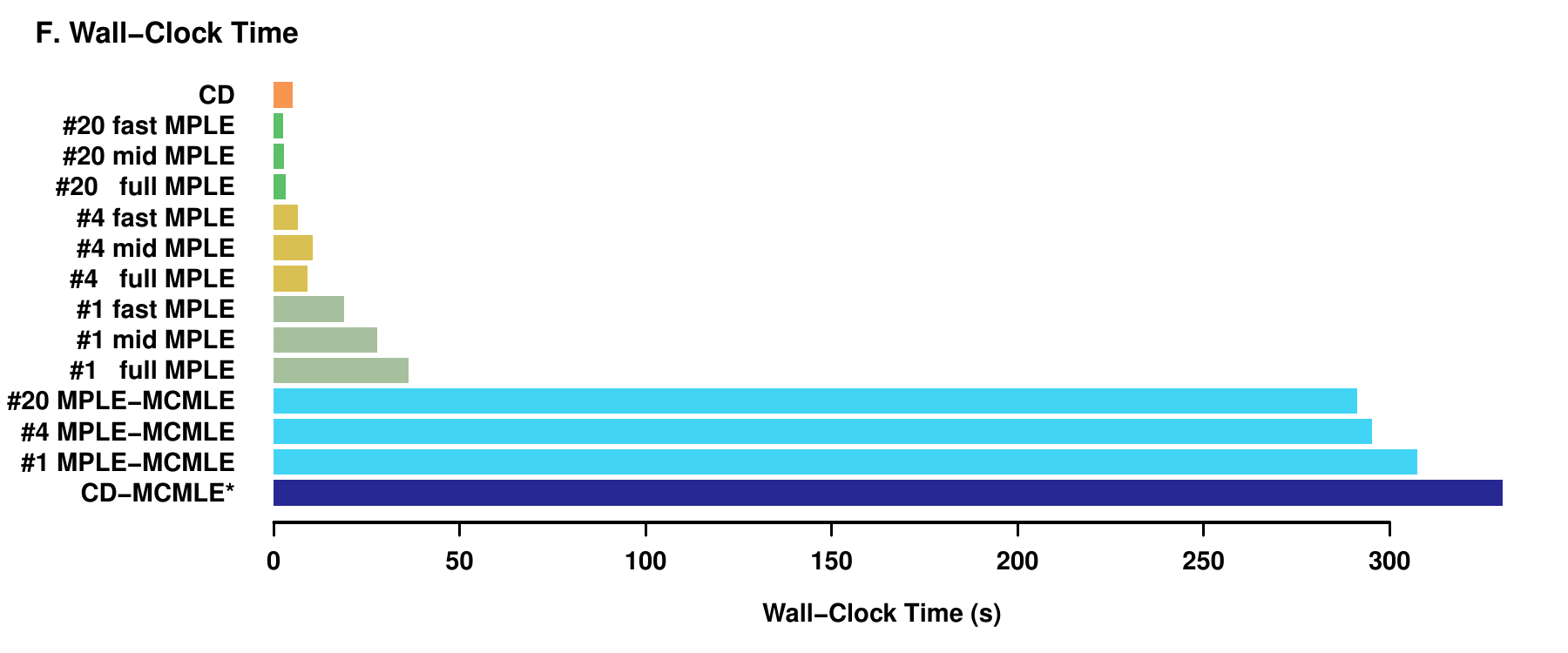}
    \end{subfigure}
    \caption{\leftline{Calibration, confidence coverage, and wall-clock time of large-variance small network}}
\label{fig:hard2}
\caption*{\emph{Note:} CD-MCMLE* results are from their final rounds with convergence. Its wall-clock time is simple summation of all rounds of computation.}
\end{figure}

Panel D in Figure~\ref{fig:hard2} shows that CD substantially overestimates the standard errors, so overly conservative uncertainty estimation is a consistent behavior for CD across all network size and edge variance structures studied. By contrast, MPLE no longer suffers from calibration difficulties for the large-variance case, producing high-quality standard error estimates that match the performance of MCMLE, seeded by either CD or MPLE. Note that for the nonzero term, CD did not return an estimate for 22 of the 500 simulated networks, meaning that the calibration of CD for the nonzero term could actually be worse; we did not rerun them, as CD's uncertainty estimates are not useful even when the algorithm did converge. Panel E in Figure~\ref{fig:hard2} reveals that CD's confidence intervals have the over-coverage issue for every covariate except the nonzero term, which suffers from a larger first-order bias. MPLE and MCMLE, both with small bias in point estimation and uncertainty estimation, offer confidence coverage very close to the 95\% benchmark. %

Lastly, Panel F in Figure~\ref{fig:hard2} shows that for the large-variance network, MCMLE is an order of magnitude slower than CD and MPLE. Subsampling and use of multiple cores are still effective ways of reducing computational time for MPLE, and MPLE with 20 processors becomes even faster than CD. We should also note that the time reported here for CD-MCMLE is the simple summation of their all rounds of wall-clock time. In reality, users need to spend more time digging into the diagnostics of MCMC after each round, and this makes CD-MCMLE even more slower than running MPLE-MCMLE.

Overall, we observe that the variance of edge values makes a substantial difference in the behavior of these estimation methods. CD generates fair point estimates in a speedy manner (albeit less accurate than its peers), but its calibration is overly conservative to the point of being unusable. The larger biases of CD estimators sometimes prevents convergence of CD-MCMLE, making MPLE a better seeding method in this scenario. MCMLE, seeded by either CD or MPLE, once converged, offers high-quality estimates at correspondingly high cost. Strikingly, however, we find that in this case MPLE generates estimators that match MCMLE in all metrics and introduces even less bias. Considering that MPLE is also an order of magnitude faster than MCMLE, it is clearly the superior method for the large-variance case. If higher-quality estimates are needed, one can increase the sample size of MPLE or follow it up with MCMLE, though the former is a much faster option.

\section{Discussion} \label{sec_disc}
Overall, this comparative simulation study reveals that the variance of the edge value makes a substantial difference in the performance of estimation methods for valued ERGMs, while network size primarily impacts computational cost. For small-variance data, all methods perform very well in point estimation, while CD greatly overestimates the uncertainty, and MPLE underestimates uncertainties of dependence terms. MCMLE seeded by either CD or MPLE offers high-quality estimates under all metrics. Wall-clock time of the methods are close to each other for small-variance small networks, but the speed advantages over MCMLE gets larger for CD and subsampled/multi-core MPLE as the network size increases.  For very large graphs (especially when large numbers of cores become available), the relative speed advantage of the MPLE can become substantial, and may be a reasonable consideration in method selection.

For large-variance data, CD fails to offer reliable uncertainty estimates; its relatively high bias compared to MPLE also makes CD-MCMLE more prone to convergence failure. Both MPLE and MCMLE (given convergence of the latter) perform well in all quality evaluations, but MPLE is an order of magnitude faster than MCMLE, and can be further sped up by subsampling and parallel computing.  Again, we observe that the speed advantages of MPLE become larger as graph size and edge variance increase.

\hl{The results suggest that for small networks with low-variance edges, MCMLE continues to serve as the estimation method of choice: it delivers high-quality point estimates and excellent calibration while still being computationally accessible in this regime. As network size increases, MCMLE becomes increasingly cumbersome, as its computational time scales up faster than other methods. When MCMLE is too slow to perform well, CD and MPLE can serve as useful tools for tasks that only require first-order point estimation, such as exploratory analysis, prediction, or generating models for network simulation. The subsampling and multi-core features of MPLE offer useful tools for fast computation of models on large networks, and the numbers of edge variables that must be sampled for strong performance are an increasingly small fraction of the total dyad count as network size grows, further enhancing its computational advantages.  Although MPLE calibration is certainly good enough to be useful (especially for independence terms), we clearly see the tendency towards overconfidence in dependence terms found in binary ERGM studies, and nominal confidence intervals for these terms are likely to be too small.  Analyses relying on coverage for such terms should be regarded as heuristic (though the maximum extent of miscalibration seen here may provide some guidance with respect to the degree of error that could be present).}

For networks with large edge variance, MPLE could be the go-to method, yielding accurate estimators with good calibration. Besides subsampling and parallel computing choices, our implementation of edgewise support truncation further offers MPLE an edge in speed without compromising its estimation quality (as edgewise and nonedgewise truncation of MPLE offers commensurate results in quality metrics for large variance data). This enables the use of MPLE to estimate valued ERGMs that were previously blocked by computational barriers, such as high-dimensional models on large networks with high edge variance. It also offers a flexible framework to make trade-offs between estimation quality and computational time, by tuning the sample size of the edge variables and the structure of the edge support truncation. While MPLE has been seen as a sub-optimal choice for binary ERGMs \citep{van_duijn_framework_2009}, this comparative simulation study reveals an area where it can be the effective approach. 

Our experiments also offer insights about MCMC-based estimation methods. This paper shows that MCMLE for valued ERGMs depends on high-quality starting values, especially for large-variance networks. Both CD and MPLE are useful tools for MCMLE seeding, but the relatively larger biases of CD for large-variance data makes CD-MCMLE more fragile. One potential reason for this fragility is the difficulty of CD in reproducing the dichotomized density of the networks (i.e. the proportion of nonzero edges). With larger ranges of edge values, the toggling of values between zero and one become more unlikely, and the difficulty of matching the target density increases. (This is, of course, a special case of zero-inflation, a common phenomenon in count data models beyond network settings.) This is reflected by the phenomenon that the sufficient statistic consistently observed to  fail in CD-MCMLE was the nonzero edge count in the large-variance experiment. The issue can be worsened by the mismatch of the MCMC algorithm design and typical properties of valued networks. Compared to their binary counterparts, valued social networks are frequently denser (in the dichotomized density). However, MCMC algorithms in existing software are optimized for sparse, unvalued social networks. We observed improvement in computational time when switching from an MCMC proposal that favors toggling nonzero edges (so-called TNT, or more accurately ``tie-random dyad'') to a random proposal, which is the one that offers the highest likelihood of toggling empty edges among existing algorithms. To improve the performance of MCMC-based methods for valued ERGMs, future research could consider experimenting with MCMC algorithms that pay more attention to the toggling between value zero and one, such as proposals that favors toggling zero-value edges.

As with any simulation study, one trades off the ``realism'' of performance on a realistic case against some degree of generality. Although we vary the network size and edge value variance to emulate different application settings, we cannot rule out the possibility that some methods studied here may perform better or worse under other conditions.  We encourage future research using simulation studies based on different use cases and model specifications, including non-Poissonian reference measures. Given that we find that the edge value variance plays an important role in influencing the performance of valued ERGM estimation, it would be of interest to experiment with more fine-grained classification of the scale of edge variance, in search of an empirical rule of thumb for when MCMLE or MPLE would be the better choice.

Our study also suggests the continuing relevance of the MPLE to ERGM methodology. Our implementation of MPLE for valued ERGMs enables estimation for large-variance data in feasible time and with high-quality results. With good overall accuracy, high speed, and flexible tunability, MPLE would be an excellent general use estimation method for valued ERGMs if its calibration could be improved for small-variance networks. Our findings suggest the value of work on methods that may help further improve calibration of MPLE in the count-valued case; such advances may build on methods that shown to help calibration for binary ERGMs, such as bootstrap resampling \citep{desmarais_statistical_2012-2,schmid_exponential_2017} and regularization \citep{van_duijn_framework_2009}. 

\hl{ Lastly, we should emphasize that it is important to perform model evaluation after estimation for generative network models like ERGMs.  While this should be a standard procedure regardless of the estimation method, it is an especially important reminder as the field observes the revival of non-simulation and local-simulation methods such as MPLE and CD, thanks to emerging methodological innovations and new data structures. These methods are less prone to convergence failures, which can have the hidden liability of making it harder to spot poorly-behaved models (an issue encountered in the early use of the MPLE before the availability of simulation-based evaluation, as discussed by \citet{snijders_markov_2002}). We recommend that researchers evaluate model adequacy by simulating networks from the fitted model and comparing their key network features and specified sufficient statistics with their observed counterparts, e.g. following the procedure of \cite{hunter_goodness_2008}, where feasible. Fortunately, simulation-based evaluation is computationally much cheaper than simulation-based estimation, as the former only requires simulation from one model while the latter needs to explore a set of models in the parameter space; thus, even when MCMLE is computationally prohibitive for evaluation, MCMC adequacy checks (a.k.a. goodness-of-fit checks) are often feasible. For sufficiently large, high-variance systems in which even this is infeasible, alternative checks are needed. Although this regime remains an open problem, conditional simulation using e.g. Held-Out Predictive Evaluation \citep{wang.et.al:sn:2016,yin.et.al:tr:2019a} may be one useful approach, provided that enough dependence-graph adjacent edge variables are held out simultaneously to permit detection of degeneracy.  Some work has been done on bounding techniques for dichotomous networks that can in some cases rule out degeneracy without resorting to simulation \citep{butts:sm:2011b}; it is unclear whether similar techniques can be developed for count-valued networks, but if so such methods could prove useful where simulation is impractical.  In general, evaluation for networks that are too large for complete simulation (in the count-data case or otherwise) is an important frontier for future work.}

\section{Conclusion} \label{sec_conc}

ERGMs, especially for valued networks, can be computationally expensive to estimate. In searching for a fast and reliable computational method, we implemented MPLE for count-valued ERGMs, and performed a comparative simulation study using three methods: CD, MCMLE, and MPLE. We found that the variance of edge values is critical in determining the performance of computational methods for valued ERGMs, while the network size mainly influences their relative merit in computational efficiency. For small-variance networks, point estimates are easy to acquire using whichever method, while CD greatly overestimates uncertainties and MPLE underestimates them for dependence terms. All methods have similar wall-clock time. For large-variance networks, both MPLE and MCMLE offer strong performance for estimating both coefficient and uncertainties, although MPLE is an order of magnitude faster than MCMLE.

On the basis of this study, we recommend that researchers pay attention to the variance of edge values in choosing computational methods. For small-variance data, MCMLE should be the default method where feasible, although CD is useful for point estimations; MPLE is suited for large networks and high-dimensional models, especially with a large number of available processors, but caveats should be given for interpreting its standard errors for dependence terms. For large-variance networks, MPLE is a solid method, and researchers can design the size of edge sample and the structure of edge support truncation based on the computational resources at hand and the requirements of estimation quality. Our experiments also demonstrate that both CD and MPLE are useful tools for MCMLE seeding, although CD is better for simpler cases with its speed advantage and MPLE is better able to offer high-quality seeds across all scenarios. 

In summary, with insights about the behaviors of each method under different network sizes and edge variances, this paper offers a guideline for choosing and tuning computational methods for valued ERGM estimation. The implementation of a flexible subsampled parallelizable MPLE framework is demonstrated to be a powerful tool; we envision it to empower researchers with large-variance big network data and high-dimensional model design, freeing them from the need to employ data-reduction and model-simplification compromises because of computational constraints.

\section*{Funding Information}
This work was supported by National Science Foundation award SES-1826589 and NIH award 1R01GM144964-01.

\section*{Acknowledgements}
We thank Katherine Faust, David Schaefer, and attendees of the Social Networks Research Group at UC Irvine for their comments and suggestions.

\section*{Appendix A. contrastive divergence with different parameters} \label{sec_CD}
Results in Section~\ref{sec_results} showed that despite its time efficiency, CD has two limitations. First, its bias is larger than subsampled MPLE, making it a suboptimal seeding method for MCMLE (especially when the edge variance is large). Second, its calibration of uncertainty is overly conservative, making it an uninformative method for second moment estimates. This leads to the question of whether one could tweak its tuning parameters to trade its time efficiency for less biased and better calibrated estimations. In this regard, we study the quality of CD estimators when we vary CD's major parameters: steps and multiplicity. ``Steps'' determines the number of Metropolis-Hastings steps, and ``multiplicity'' determines the number of proposal for each step. The default setting in \texttt{ergm.count} package for CD is 8 steps and 1 multiplicity, which was the setting reported in Section~\ref{sec_results}. Here we compare that with different combinations of modified tuning parameters.

\begin{table}[ht]
\renewcommand{\thetable}{A\arabic{table}}  %
\setcounter{table}{0}
\caption{\leftline{Bias and time of CD vs. MPLE}}
\label{tab:cdbias}
\begin{tabular}{lrrrrrrr}
\hline
 & \multicolumn{6}{c}{Contrastive Divergence} & MPLE \\ \hline
Steps & 8 & 80 & 8 & 80 & 800 & 8000 &  \\ 
  Multiplicity & 1 & 1 & 10 & 10 & 1 & 1 & \\  \hline 
  \multicolumn{8}{l}{\emph{Small-Variance Small Network}}\\
  Bias(Mutual) (\%) & 0.07 & 0.64 & 1.07 & 1.31 & 0.64 & 0.53 & 0.66 \\ 
  Bias(Flow) (\%) & 1.64 & 1.38 & 0.51 & 0.68 & 2.54 & 3.53 & 0.35 \\ 
  Wall-clock time (s) & 0.60 & 1.92 & 2.01 & 21.55 & 30.77 & 342.95 & 26.14 \\ \hline
  \multicolumn{8}{l}{\emph{Small-Variance Large Network}}\\
  Bias(Mutual) (\%) & 0.93 & 0.55 & 1.66 & 0.60 & 0.34 & 0.30 & 0.12 \\ 
  Bias(Flow) (\%) & 2.35 & 1.10 & 1.98 & 1.17 & 0.89 & 0.72 & 0.15 \\  
  Wall-clock time (s) & 1.15 & 6.00 & 6.34 & 69.82 & 96.63 & 2177.46 & 769.77 \\ \hline
  \multicolumn{8}{l}{\emph{Large-Variance Small Network}}\\
    Bias(Mutual) (\%) & 0.42 & 79.94 & 87.41 & 99.4 & 139.9 & 116.13 & 0.05 \\ 
  Bias(Flow) (\%) & 3.86 & 591.9 & 469.13 & 261.75 & 97.59 & 29.45 & 1.78 \\  
  Wall-clock time (s) & 5.22 & 12.57 & 13.02 & 94.71 & 81.46 & 876.71 & 18.83 \\ \hline
\end{tabular}
\caption*{\emph{Note:} We use the chosen seeding setting for MPLE: 50\% sample sizes for large-variance and large-network data, 75\% for small-variance small network data. Wall-clock time of MPLE is from the slowest setting using one core.}
\end{table}

\begin{table}[h]
\renewcommand{\thetable}{A\arabic{table}}  %
\caption{\leftline{Calibration and time of CD vs. MCMLE}}
\label{tab:cdcalib}
\begin{tabular}{lrrrrrrr}
\hline
 & \multicolumn{6}{c}{Contrastive Divergence} & MCMLE \\ \hline
Steps & 8 & 80 & 8 & 80 & 800 & 8000 &  \\ 
  Multiplicity & 1 & 1 & 10 & 10 & 1 & 1 &  \\  \hline 
  \multicolumn{8}{l}{\emph{Small-Variance Small Network}}\\  
  Mutual & 2.11 & 1.04 & 2.73 & 1.75 & 0.22 & 0.05 & ~0.04 \\ 
  Flow & 1.88 & 0.93 & 2.57 & 1.51 & 0.19 & -0.02 & -0.03 \\ 
  Wall-clock time (s) & 0.60 & 1.92 & 2.01 & 21.55 & 30.77 & 342.95 & 30.16 \\ 
  \hline 
  \multicolumn{8}{l}{\emph{Small-Variance Large Network}}\\
    Mutual & 2.91 & 1.99 & 3.21 & 2.75 & 0.95 & 0.16 & -0.03 \\ 
  Flow & 2.76 & 1.88 & 3.10 & 2.53 & 0.98 & 0.21 & -0.09 \\ 
  Wall-clock time (s) & 1.15 & 6.00 & 6.34 & 69.82 & 96.63 & 2177.46 & 1673.59 \\ 
   \hline
  \multicolumn{8}{l}{\emph{Large-Variance Small Network}}\\
    Mutual & 2.55 & 0.97 & 1.06 & 0.36 & -0.33 & -0.56 & -0.05 \\ 
  Flow & 2.63 & 0.33 & 1.23 & 0.05 & -0.70 & -0.96 & -0.11 \\ 
  Wall-clock time (s) & 5.22 & 12.57 & 13.02 & 94.71 & 81.46 & 876.71 & 307.35 \\
  \hline 
\end{tabular}
\caption*{\leftline{\emph{Note:} MCMLE is seeded by MPLE under the specified setting in Results using one core.}}
\end{table}

Table~\ref{tab:cdbias} shows the bias and the wall-clock time for CD under different tuning parameters versus MPLE under its configuration for MCMLE seeding using a single processor. In this section, we only report performance for the dependence terms because of space limitation, but estimators for other covariates generally share similar patterns. For both the small-variance small network and the large-variance small networks, increasing either steps or multiplicity or both does not monotonically reduce the bias, while the wall-clock time increases monotonically as expected. For this reason, the default CD configuration seems to be the optimal choice as a seeding method for MCMLE. For the small-variance large network, although larger multiplicity does not bring less bias, increasing steps does brings monotonic decrease in bias estimation. However, CD's bias is always larger than that of MPLE, even when its wall-clock time surpasses MPLE's. This suggests that for small-variance large networks, one can increase steps to reduce the bias of CD, but it is not as efficient as using MPLE instead, which offers better estimators with less time used. To recap, increasing either or both tuning parameters for CD usually fails to yield less biased estimators, and even when that does, it is not as time-efficient as using MPLE.

Table~\ref{tab:cdcalib} displays the calibration for the two dependence terms and wall-clock time of CD under various tuning parameters versus the benchmark MCMLE, seeded by MPLE with a single processor. For both the small-variance small network and the small-variance large network, increasing multiplicity does not lead to better calibrated standard errors, but increasing steps is associated with monotonic improvement in calibration. Nonetheless, CD fails to offer as well-calibrated estimates as MCMLE in comparable time spans. This suggests that, for small-variance data, compared to increasing steps for CD, it is a better choice to directly use MCMLE for a time-efficient and well-calibrated second-moment measurement. For large-variance small-network data, although increasing steps and/or multiplicity in general alleviate its overestimation of uncertainty, increasing steps beyond a certain point can lead to underestimation of uncertainty. Regardless, its calibration never beats MCMLE. Overall, our experiments suggest that increasing tuning parameters beyond the default setting in \texttt{statnet} does not always improve its calibration, and when it does, it is not as time-efficient as using MCMLE for calibration.

\section*{Appendix B. Supplementary Material}
Replication data and the R source code for the MPLE can be found at: \href{https://doi.org/10.7910/DVN/BFVXZ6}{doi:10.7910/DVN/BFVXZ6}

\bibliographystyle{asr.bst}
\bibliography{zotero}

\begin{thebibliography}{}
\newcommand{\enquote}[1]{``#1''}

\bibitem[\protect\citeauthoryear{Aicher, Jacobs, and Clauset}{Aicher
  et~al.}{2014}]{aicher.et.al:jcn:2014}
Aicher, Christopher, Abigail~Z. Jacobs, and Aaron Clauset. 2014.
\newblock \enquote{{Learning latent block structure in weighted networks}.}
\newblock {\em Journal of Complex Networks\/} 3:221--248.

\bibitem[\protect\citeauthoryear{Aksoy and Yıldırım}{Aksoy and
  Yıldırım}{2020}]{aksoy_model_2020}
Aksoy, Ozan and Sinan Yıldırım. 2020.
\newblock \enquote{A model of dynamic flows: {Explaining} {Turkey}'s
  inter-provincial migration.}
\newblock preprint, SocArXiv.

\bibitem[\protect\citeauthoryear{Altman and Royston}{Altman and
  Royston}{2006}]{altman2006cost}
Altman, Douglas~G and Patrick Royston. 2006.
\newblock \enquote{The cost of dichotomising continuous variables.}
\newblock {\em Bmj\/} 332:1080.

\bibitem[\protect\citeauthoryear{An}{An}{2016}]{an_fitting_2016}
An, Weihua. 2016.
\newblock \enquote{Fitting {ERGMs} on big networks.}
\newblock {\em Social Science Research\/} 59:107--119.

\bibitem[\protect\citeauthoryear{Anderson, Wasserman, and Crouch}{Anderson
  et~al.}{1999}]{c.anderson.et.al:sn:1999}
Anderson, C., S.~Wasserman, and B.~Crouch. 1999.
\newblock \enquote{A p* Primer: Logit Models for Social Networks.}
\newblock {\em Social Networks\/} 21:37--66.

\bibitem[\protect\citeauthoryear{Asuncion, Liu, A.~Ihler, and Smyth}{Asuncion
  et~al.}{2010}]{ascuncion.et.al:icml:2010}
Asuncion, A., Q.~Liu, P~A.~Ihler, and Smyth. 2010.
\newblock \enquote{Particle Filtered {MCMC-MLE} with Connections to Contrastive
  Divergence.}
\newblock In {\em International Conference on Machine Learning (ICML)\/}.

\bibitem[\protect\citeauthoryear{Bernard, Killworth, , and Sailer}{Bernard
  et~al.}{1979}]{bernard.et.al:sn:1979}
Bernard, H.~Russell, Peter Killworth, , and Lee Sailer. 1979.
\newblock \enquote{Informant Accuracy in Social Networks {IV}: A Comparison of
  Clique-Level Structure in Behavioral and Cognitive Network Data.}
\newblock {\em Social Networks\/} 2:191--218.

\bibitem[\protect\citeauthoryear{Besag}{Besag}{1974}]{besag_spatial_1974}
Besag, Julian. 1974.
\newblock \enquote{Spatial {Interaction} and the {Statistical} {Analysis} of
  {Lattice} {Systems}.}
\newblock {\em Journal of the Royal Statistical Society: Series B
  (Methodological)\/} 36:192--225.

\bibitem[\protect\citeauthoryear{Bhamidi, Bresler, and Sly}{Bhamidi
  et~al.}{2011}]{bhamidi.et.al:aap:2011}
Bhamidi, Shankar, Guy Bresler, and Allan Sly. 2011.
\newblock \enquote{Mixing Time of Exponential Random Graphs.}
\newblock {\em The Annals of Applied Probability\/} pp. 2146--2170.

\bibitem[\protect\citeauthoryear{Block, Stadtfeld, and Robins}{Block
  et~al.}{2022}]{block2022statistical}
Block, Per, Christoph Stadtfeld, and Garry Robins. 2022.
\newblock \enquote{A statistical model for the analysis of mobility tables as
  weighted networks with an application to faculty hiring networks.}
\newblock {\em Social Networks\/} 68:264--278.

\bibitem[\protect\citeauthoryear{Boyle, Keith~H., Vaughan, and Vaughan}{Boyle
  et~al.}{2014}]{boyle_exploring_2014}
Boyle, Paul, Halfacree Keith~H., Robinson Vaughan, and Robinson Vaughan. 2014.
\newblock {\em Exploring {Contemporary} {Migration}\/}.
\newblock Abingdon, United Kingdom: Routledge.

\bibitem[\protect\citeauthoryear{Butts}{Butts}{2008}]{butts_social_2008}
Butts, Carter~T. 2008.
\newblock \enquote{Social {Network} {Analysis} with sna.}
\newblock {\em Journal of Statistical Software\/} 24:1--51.

\bibitem[\protect\citeauthoryear{Butts}{Butts}{2011}]{butts:sm:2011b}
Butts, Carter~T. 2011.
\newblock \enquote{{B}ernoulli Graph Bounds for General Random Graphs.}
\newblock {\em Sociological Methodology\/} 41:299--345.

\bibitem[\protect\citeauthoryear{Butts}{Butts}{2019}]{butts:jms:2019}
Butts, Carter~T. 2019.
\newblock \enquote{A Dynamic Process Interpretation of the Sparse {ERGM}
  Reference Model.}
\newblock {\em Journal of Mathematical Sociology\/} 43:40--57.

\bibitem[\protect\citeauthoryear{Butts}{Butts}{2020}]{butts:jms:2020a}
Butts, Carter~T. 2020.
\newblock \enquote{A Dynamic Process Reference Model for Sparse Networks with
  Reciprocity.}
\newblock {\em Journal of Mathematical Sociology\/} .

\bibitem[\protect\citeauthoryear{Butts, Petrescu-Prahova, and Cross}{Butts
  et~al.}{2007}]{butts.et.al:jms:2007}
Butts, Carter~T., Miruna Petrescu-Prahova, and B.~Remy Cross. 2007.
\newblock \enquote{Responder Communication Networks in the {W}orld {T}rade
  {C}enter {D}isaster: Implications for Modeling of Communication within
  Emergency Settings.}
\newblock {\em Journal of Mathematical Sociology\/} 31:121--147.

\bibitem[\protect\citeauthoryear{Cranmer and Desmarais}{Cranmer and
  Desmarais}{2011}]{cranmer_inferential_2011}
Cranmer, Skyler~J. and Bruce~A. Desmarais. 2011.
\newblock \enquote{Inferential {Network} {Analysis} with {Exponential} {Random}
  {Graph} {Models}.}
\newblock {\em Political Analysis\/} 19:66--86.

\bibitem[\protect\citeauthoryear{Dekker, Krackhardt, and Snijders}{Dekker
  et~al.}{2007}]{dekker.et.al:p:2007}
Dekker, David, David Krackhardt, and Tom A.~B. Snijders. 2007.
\newblock \enquote{Sensitivity of {MRQAP} Tests to Collinearity and
  Autocorrelation Conditions.}
\newblock {\em Psychometrika\/} 72:563--581.

\bibitem[\protect\citeauthoryear{Desmarais and Cranmer}{Desmarais and
  Cranmer}{2012a}]{desmarais_statistical_2012}
Desmarais, Bruce~A. and Skyler~J. Cranmer. 2012a.
\newblock \enquote{Statistical {Inference} for {Valued}-{Edge} {Networks}:
  {The} {Generalized} {Exponential} {Random} {Graph} {Model}.}
\newblock {\em PLoS ONE\/} 7:e30136.

\bibitem[\protect\citeauthoryear{Desmarais and Cranmer}{Desmarais and
  Cranmer}{2012b}]{desmarais_statistical_2012-2}
Desmarais, Bruce~A. and Skyler~J. Cranmer. 2012b.
\newblock \enquote{Statistical mechanics of networks: {Estimation} and
  uncertainty.}
\newblock {\em Physica A: Statistical Mechanics and its Applications\/}
  391:1865--1876.

\bibitem[\protect\citeauthoryear{Drabek, Tamminga, Kilijanek, and Adams}{Drabek
  et~al.}{1981}]{drabek.et.al:bk:1981}
Drabek, Thomas~E., Harriet~L. Tamminga, Thomas~S. Kilijanek, and Christopher~R.
  Adams. 1981.
\newblock {\em Managing Multiorganizational Emergency Responses: Emergent
  Search and Rescue Networks in Natural Disaster and Remote Area Settings\/}.
\newblock Number Monograph 33 in Program on Technology, Environment, and Man.
  Boulder, CO: Institute of Behavioral Sciences, University of Colorado.

\bibitem[\protect\citeauthoryear{Eddelbuettel, Fran{\c{c}}ois, Allaire, Ushey,
  Kou, Russel, Chambers, and Bates}{Eddelbuettel
  et~al.}{2011}]{eddelbuettel2011rcpp}
Eddelbuettel, Dirk, Romain Fran{\c{c}}ois, J~Allaire, Kevin Ushey, Qiang Kou,
  N~Russel, John Chambers, and D~Bates. 2011.
\newblock \enquote{Rcpp: Seamless R and C++ integration.}
\newblock {\em Journal of Statistical Software\/} 40:1--18.

\bibitem[\protect\citeauthoryear{Faust}{Faust}{2011}]{faust2011animal}
Faust, Katherine. 2011.
\newblock \enquote{Animal social networks.}
\newblock {\em The SAGE Handbook of Social Network Analysis\/} 148:166.

\bibitem[\protect\citeauthoryear{Fowler}{Fowler}{2006}]{fowler2006connecting}
Fowler, James~H. 2006.
\newblock \enquote{Connecting the Congress: A study of cosponsorship networks.}
\newblock {\em Political Analysis\/} 14:456--487.

\bibitem[\protect\citeauthoryear{Geyer and Thompson}{Geyer and
  Thompson}{1992}]{geyer_constrained_1992}
Geyer, Charles~J. and Elizabeth~A. Thompson. 1992.
\newblock \enquote{Constrained {Monte} {Carlo} {Maximum} {Likelihood} for
  {Dependent} {Data}.}
\newblock {\em Journal of the Royal Statistical Society: Series B
  (Methodological)\/} 54:657--683.

\bibitem[\protect\citeauthoryear{Granovetter}{Granovetter}{1973}]{granovetter_strength_1973}
Granovetter, Mark~S. 1973.
\newblock \enquote{The {Strength} of {Weak} {Ties}.}
\newblock {\em American Journal of Sociology\/} 78:1360--1380.

\bibitem[\protect\citeauthoryear{Handcock}{Handcock}{2003}]{handcock:ch:2003}
Handcock, Mark~S. 2003.
\newblock \enquote{Statistical Models for Social Networks: Inference and
  Degeneracy.}
\newblock In {\em Dynamic Social Network Modeling and Analysis\/}, edited by
  Ron Breiger, Kathleen~M. Carley, and Philippa Pattison, pp. 229--240.
  Washington, DC: National Academies Press.

\bibitem[\protect\citeauthoryear{Handcock, Hunter, Butts, Goodreau, and
  Morris}{Handcock et~al.}{2008}]{handcock_statnet:_2008}
Handcock, Mark~S., David~R. Hunter, Carter~T. Butts, Steven~M. Goodreau, and
  Martina Morris. 2008.
\newblock \enquote{statnet: {Software} {Tools} for the {Representation},
  {Visualization}, {Analysis} and {Simulation} of {Network} {Data}.}
\newblock {\em Journal of Statistical Software\/} 24:1548--7660.

\bibitem[\protect\citeauthoryear{Hinton}{Hinton}{2002}]{hinton_training_2002}
Hinton, Geoffrey~E. 2002.
\newblock \enquote{Training {Products} of {Experts} by {Minimizing}
  {Contrastive} {Divergence}.}
\newblock {\em Neural Computation\/} 14:1771--1800.

\bibitem[\protect\citeauthoryear{Hoff, Raftery, and Handcock}{Hoff
  et~al.}{2002}]{hoff.et.al:jasa:2002}
Hoff, Peter~D., Adrian~E. Raftery, and Mark~S. Handcock. 2002.
\newblock \enquote{Latent Space Approaches to Social Network Analysis.}
\newblock {\em Journal of the American Statistical Association\/}
  97:1090--1098.

\bibitem[\protect\citeauthoryear{Huang and Butts}{Huang and
  Butts}{2022}]{huang2022rooted}
Huang, Peng and Carter~T Butts. 2022.
\newblock \enquote{Rooted America: Immobility and Segregation of the
  Inter-county Migration Networks.}
\newblock {\em arXiv preprint arXiv:2205.02347\/} .

\bibitem[\protect\citeauthoryear{Hummel, Hunter, and Handcock}{Hummel
  et~al.}{2012}]{hummel_improving_2012}
Hummel, Ruth~M., David~R. Hunter, and Mark~S. Handcock. 2012.
\newblock \enquote{Improving {Simulation}-{Based} {Algorithms} for {Fitting}
  {ERGMs}.}
\newblock {\em Journal of Computational and Graphical Statistics\/}
  21:920--939.

\bibitem[\protect\citeauthoryear{Hunter, Goodreau, and Handcock}{Hunter
  et~al.}{2008a}]{hunter_goodness_2008}
Hunter, David~R., Steven~M. Goodreau, and Mark~S. Handcock. 2008a.
\newblock \enquote{Goodness of {Fit} of {Social} {Network} {Models}.}
\newblock {\em Journal of the American Statistical Association\/} 103:248--258.

\bibitem[\protect\citeauthoryear{Hunter, Handcock, Butts, Goodreau, and
  Morris}{Hunter et~al.}{2008b}]{hunter_ergm_2008}
Hunter, David~R., Mark~S. Handcock, Carter~T. Butts, Steven~M. Goodreau, and
  Martina Morris. 2008b.
\newblock \enquote{ergm: {A} {Package} to {Fit}, {Simulate} and {Diagnose}
  {Exponential}-{Family} {Models} for {Networks}.}
\newblock {\em Journal of Statistical Software\/} 24:nihpa54860.

\bibitem[\protect\citeauthoryear{Hunter, Krivitsky, and Schweinberger}{Hunter
  et~al.}{2012}]{hunter2012computational}
Hunter, David~R, Pavel~N Krivitsky, and Michael Schweinberger. 2012.
\newblock \enquote{Computational statistical methods for social network
  models.}
\newblock {\em Journal of Computational and Graphical Statistics\/}
  21:856--882.

\bibitem[\protect\citeauthoryear{Hyv{\"a}rinen}{Hyv{\"a}rinen}{2006}]{hyvarinen2006consistency}
Hyv{\"a}rinen, Aapo. 2006.
\newblock \enquote{Consistency of pseudolikelihood estimation of fully visible
  Boltzmann machines.}
\newblock {\em Neural Computation\/} 18:2283--2292.

\bibitem[\protect\citeauthoryear{Krackhardt}{Krackhardt}{1988}]{krackhardt:sn:1988}
Krackhardt, David. 1988.
\newblock \enquote{Predicting with Networks: Nonparametric Multiple Regression
  Analyses of Dyadic Data.}
\newblock {\em Social Networks\/} 10:359--382.

\bibitem[\protect\citeauthoryear{Krivitsky}{Krivitsky}{2012}]{krivitsky_exponential-family_2012}
Krivitsky, Pavel~N. 2012.
\newblock \enquote{Exponential-family random graph models for valued networks.}
\newblock {\em Electronic Journal of Statistics\/} 6:1100--1128.

\bibitem[\protect\citeauthoryear{Krivitsky}{Krivitsky}{2017}]{krivitsky_using_2017}
Krivitsky, Pavel~N. 2017.
\newblock \enquote{Using contrastive divergence to seed {Monte} {Carlo} {MLE}
  for exponential-family random graph models.}
\newblock {\em Computational Statistics \& Data Analysis\/} 107:149--161.

\bibitem[\protect\citeauthoryear{Krivitsky and Butts}{Krivitsky and
  Butts}{2017}]{krivitsky_exponential-family_2017}
Krivitsky, Pavel~N. and Carter~T. Butts. 2017.
\newblock \enquote{Exponential-family {Random} {Graph} {Models} for
  {Rank}-order {Relational} {Data}.}
\newblock {\em Sociological Methodology\/} 47:68--112.

\bibitem[\protect\citeauthoryear{Krivitsky, Handcock, and Hunter}{Krivitsky
  et~al.}{2012}]{krivitsky_package_2012}
Krivitsky, Pavel~N., Mark~S. Handcock, and David~R. Hunter. 2012.
\newblock \enquote{Package ‘ergm.count’.}
\newblock {\em Journal of Statistics\/} 6:1100--1128.

\bibitem[\protect\citeauthoryear{Krivitsky, Hunter, Morris, and
  Klumb}{Krivitsky et~al.}{2022}]{krivitsky2022ergm}
Krivitsky, Pavel~N, David~R Hunter, Martina Morris, and Chad Klumb. 2022.
\newblock \enquote{ergm 4: Computational Improvements.}
\newblock {\em arXiv preprint arXiv:2203.08198\/} .

\bibitem[\protect\citeauthoryear{Leal}{Leal}{2021}]{leal_network_2021}
Leal, Diego~F. 2021.
\newblock \enquote{Network {Inequalities} and {International} {Migration} in
  the {Americas}.}
\newblock {\em American Journal of Sociology\/} 126:1067--1126.

\bibitem[\protect\citeauthoryear{Lubbers and Snijders}{Lubbers and
  Snijders}{2007}]{lubbers_comparison_2007}
Lubbers, Miranda~J. and Tom A.~B. Snijders. 2007.
\newblock \enquote{A comparison of various approaches to the exponential random
  graph model: {A} reanalysis of 102 student networks in school classes.}
\newblock {\em Social Networks\/} 29:489--507.

\bibitem[\protect\citeauthoryear{McMillan}{McMillan}{2022}]{mcmillan2022worth}
McMillan, Cassie. 2022.
\newblock \enquote{Worth the weight: Conceptualizing and measuring strong
  versus weak tie homophily.}
\newblock {\em Social Networks\/} 68:139--147.

\bibitem[\protect\citeauthoryear{Mele}{Mele}{2017}]{mele2017structural}
Mele, Angelo. 2017.
\newblock \enquote{A structural model of dense network formation.}
\newblock {\em Econometrica\/} 85:825--850.

\bibitem[\protect\citeauthoryear{Morris, Handcock, and Hunter}{Morris
  et~al.}{2008}]{morris_specification_2008}
Morris, Martina, Mark~S. Handcock, and David~R. Hunter. 2008.
\newblock \enquote{Specification of {Exponential}-{Family} {Random} {Graph}
  {Models}: {Terms} and {Computational} {Aspects}.}
\newblock {\em Journal of statistical software\/} 24:1548--7660.

\bibitem[\protect\citeauthoryear{Nowicki and Snijders}{Nowicki and
  Snijders}{2001}]{nowicki.snijders:jasa:2001}
Nowicki, Krzysztof and Tom A.~B. Snijders. 2001.
\newblock \enquote{Estimation and Prediction for Stochastic Blockstructures.}
\newblock {\em Journal of the American Statistical Association\/}
  96:1077--1087.

\bibitem[\protect\citeauthoryear{Robins, Pattison, and Wasserman}{Robins
  et~al.}{1999}]{robins.et.al:p:1999}
Robins, Garry~L., Philippa~E. Pattison, and Stanley Wasserman. 1999.
\newblock \enquote{Logit Models and Logistic Regressions for Social Networks,
  III. Valued Relations.}
\newblock {\em Psychometrika\/} 64:371--394.

\bibitem[\protect\citeauthoryear{Schmid and Desmarais}{Schmid and
  Desmarais}{2017}]{schmid_exponential_2017}
Schmid, Christian~S. and Bruce~A. Desmarais. 2017.
\newblock \enquote{Exponential random graph models with big networks: {Maximum}
  pseudolikelihood estimation and the parametric bootstrap.}
\newblock In {\em 2017 {IEEE} {International} {Conference} on {Big} {Data}
  ({Big} {Data})\/}, pp. 116--121.

\bibitem[\protect\citeauthoryear{Simpson, Bowman, and Laurienti}{Simpson
  et~al.}{2013}]{simpson_analyzing_2013}
Simpson, Sean~L., F.~DuBois Bowman, and Paul~J. Laurienti. 2013.
\newblock \enquote{Analyzing complex functional brain networks: {Fusing}
  statistics and network science to understand the brain.}
\newblock {\em Statistics Surveys\/} 7:1--36.

\bibitem[\protect\citeauthoryear{Snijders}{Snijders}{2002}]{snijders_markov_2002}
Snijders, Tom A~B. 2002.
\newblock \enquote{Markov {Chain} {Monte} {Carlo} {Estimation} of {Exponential}
  {Random} {Graph} {Models}.}
\newblock {\em Journal of Social Structure\/} 3:1--40.

\bibitem[\protect\citeauthoryear{Strauss and Ikeda}{Strauss and
  Ikeda}{1990}]{strauss_pseudolikelihood_1990}
Strauss, David and Michael Ikeda. 1990.
\newblock \enquote{Pseudolikelihood {Estimation} for {Social} {Networks}.}
\newblock {\em Journal of the American Statistical Association\/} 85:204--212.

\bibitem[\protect\citeauthoryear{Tan and Friel}{Tan and
  Friel}{2020}]{tan2020bayesian}
Tan, Linda~SL and Nial Friel. 2020.
\newblock \enquote{Bayesian variational inference for exponential random graph
  models.}
\newblock {\em Journal of Computational and Graphical Statistics\/}
  29:910--928.

\bibitem[\protect\citeauthoryear{Ulibarri and Scott}{Ulibarri and
  Scott}{2017}]{ulibarri_linking_2017}
Ulibarri, Nicola and Tyler~A. Scott. 2017.
\newblock \enquote{Linking {Network} {Structure} to {Collaborative}
  {Governance}.}
\newblock {\em Journal of Public Administration Research and Theory\/}
  27:163--181.

\bibitem[\protect\citeauthoryear{{U.S. Census Bureau}}{{U.S. Census
  Bureau}}{2018}]{ACSdata}
{U.S. Census Bureau}. 2018.
\newblock \enquote{{County-to-County Migration Flows: 2011-2015 ACS}.}
\newblock
  https://www.census.gov/data/tables/2015/demo/geographic-mobility/county-to-county-migration-2011-2015.html.
\newblock Accessed: 1/15/2019.

\bibitem[\protect\citeauthoryear{van Duijn, Gile, and Handcock}{van Duijn
  et~al.}{2009}]{van_duijn_framework_2009}
van Duijn, Marijtje A.~J., Krista~J. Gile, and Mark~S. Handcock. 2009.
\newblock \enquote{A framework for the comparison of maximum pseudo-likelihood
  and maximum likelihood estimation of exponential family random graph models.}
\newblock {\em Social Networks\/} 31:52--62.

\bibitem[\protect\citeauthoryear{{Vega Yon}, Slaughter, and {de la Haye}}{{Vega
  Yon} et~al.}{2021}]{von.et.al:sn:2021}
{Vega Yon}, George~G., Andrew Slaughter, and Kayla {de la Haye}. 2021.
\newblock \enquote{Exponential random graph models for little networks.}
\newblock {\em Social Networks\/} 64:225--238.

\bibitem[\protect\citeauthoryear{Vu, Hunter, and Schweinberger}{Vu
  et~al.}{2013}]{vu.et.al:aas:2013}
Vu, Duy~Q., David~R. Hunter, and Michael Schweinberger. 2013.
\newblock \enquote{{Model-based clustering of large networks}.}
\newblock {\em The Annals of Applied Statistics\/} 7:1010 -- 1039.

\bibitem[\protect\citeauthoryear{Wainwright, Jordan, et~al.}{Wainwright
  et~al.}{2008}]{wainwright2008graphical}
Wainwright, Martin~J, Michael~I Jordan, et~al. 2008.
\newblock \enquote{Graphical models, exponential families, and variational
  inference.}
\newblock {\em Foundations and Trends{\textregistered} in Machine Learning\/}
  1:1--305.

\bibitem[\protect\citeauthoryear{Wang, Butts, Hipp, Jose, and Lakon}{Wang
  et~al.}{2016}]{wang.et.al:sn:2016}
Wang, Cheng, Carter~T. Butts, John~R. Hipp, Rupa Jose, and Cynthia~M. Lakon.
  2016.
\newblock \enquote{Multiple Imputation for Missing Edge Data: A Predictive
  Evaluation Method with Application to {A}dd {H}ealth.}
\newblock {\em Social Networks\/} 45:89--98.

\bibitem[\protect\citeauthoryear{Wang, Robins, and Pattison}{Wang
  et~al.}{2009}]{wang_pnet_2009}
Wang, Peng, Garry Robins, and Philippa Pattison. 2009.
\newblock \enquote{{PNet}: {Program} for the {Simulation} and {Estimation} of
  {Exponential} {Random} {Graph} (p*) {Models}.}
\newblock {\em The University of Melbourne\/} .

\bibitem[\protect\citeauthoryear{Ward, Ahlquist, and Rozenas}{Ward
  et~al.}{2013}]{ward_gravitys_2013}
Ward, Michael~D., John~S. Ahlquist, and Arturas Rozenas. 2013.
\newblock \enquote{Gravity's {Rainbow}: {A} dynamic latent space model for the
  world trade network.}
\newblock {\em Network Science\/} 1:95--118.

\bibitem[\protect\citeauthoryear{Windzio}{Windzio}{2018}]{windzio_network_2018}
Windzio, Michael. 2018.
\newblock \enquote{The network of global migration 1990–2013.}
\newblock {\em Social Networks\/} 53:20--29.

\bibitem[\protect\citeauthoryear{Windzio, Teney, and Lenkewitz}{Windzio
  et~al.}{2019}]{windzio_network_2019}
Windzio, Michael, Céline Teney, and Sven Lenkewitz. 2019.
\newblock \enquote{A network analysis of intra-{EU} migration flows: how
  regulatory policies, economic inequalities and the network-topology shape the
  intra-{EU} migration space.}
\newblock {\em Journal of Ethnic and Migration Studies\/} pp. 1--19.

\bibitem[\protect\citeauthoryear{Yin, Phillips, and Butts}{Yin
  et~al.}{2019}]{yin.et.al:tr:2019a}
Yin, Fan, Nolan~E. Phillips, and Carter~T. Butts. 2019.
\newblock \enquote{Selection of Exponential-Family Random Graph Models via
  {H}eld-{O}ut {P}redictive {E}valuation ({HOPE}).}
\newblock arXiv:1908.05873.

\bibitem[\protect\citeauthoryear{Zipf}{Zipf}{1946}]{zipf_p1_1946}
Zipf, George~Kingsley. 1946.
\newblock \enquote{The {P1} {P2}/{D} {Hypothesis}: {On} the {Intercity}
  {Movement} of {Persons}.}
\newblock {\em American Sociological Review\/} 11:677--686.

\end{thebibliography}

\end{document}